\documentclass[journal=jacsat,manuscript=article]{achemso}

\usepackage[version=3]{mhchem} % Formula subscripts using \ce{}
\usepackage{xcolor}
\usepackage{amsmath}
\usepackage{siunitx}
\usepackage{soul}
\usepackage{color,soul}
\usepackage{overpic}
\definecolor{darkBlue}{RGB}{0,0,130}
\definecolor{darkGreen}{RGB}{13, 59, 2}

\mciteErrorOnUnknownfalse

\let\vec\mathbf

\author{Solana Di Pino}
\affiliation{Departamento de Qu\'imica Inorg\'anica, Anal\'itica
y Qu\'imica F\'isica/INQUIMAE, Facultad de Ciencias Exactas
y Naturales, Universidad de Buenos Aires, Ciudad Universitaria,
Buenos Aires (C1428EHA) Argentina\\}

\author{Edward Danquah Donkor}
\affiliation{International Centre for Theoretical Physics, Strada Costiera 11, 34151 Trieste, Italy}
\altaffiliation{Scuola Internazionale Superiore di Studi Avanzati (SISSA),
34136 Trieste, Italy}

\author{Veronica M. S\'anchez}
\affiliation{Departamento de Qu\'imica Inorg\'anica, Anal\'itica
y Qu\'imica F\'isica/INQUIMAE, Facultad de Ciencias Exactas
y Naturales, Universidad de Buenos Aires, Ciudad Universitaria,
Buenos Aires (C1428EHA) Argentina\\}

\author{Alex Rodriguez}
\affiliation{International Centre for Theoretical Physics, Strada Costiera 11, 34151 Trieste, Italy}
\affiliation{University of Trieste, Trieste, Italy.}

\author{Giuseppe Cassone}
\affiliation{Institute for Chemical-Physical Processes, National Research Council (CNR-IPCF), Viale Stagno d'Alcontres 37, 98158 Messina, Italy}

\author{Damian Scherlis}
\affiliation{Departamento de Qu\'imica Inorg\'anica, Anal\'itica
y Qu\'imica F\'isica/INQUIMAE, Facultad de Ciencias Exactas
y Naturales, Universidad de Buenos Aires, Ciudad Universitaria,
Buenos Aires (C1428EHA) Argentina\\}

\author{Ali Hassanali}
\affiliation{International Centre for Theoretical Physics, Strada Costiera 11, 34151 Trieste, Italy}
\email{ahassana@ictp.it}
\title[]
  {\emph{ZundEig}: The Structure of the Proton in Liquid Water From Unsupervised Learning
  }

\abbreviations{IR,NMR,UV}

\begin{document}

%%%%%%%%%%%%%%%%%%%%%%%%%%%%%%%%%%%%%%%%%%%%%%%%%%%%%%%%%%%%%%%%%%%%%
%% The "tocentry" environment can be used to create an entry for the
%% graphical table of contents. It is given here as some journals
%% require that it is printed as part of the abstract page. It will
%% be automatically moved as appropriate.
%%%%%%%%%%%%%%%%%%%%%%%%%%%%%%%%%%%%%%%%%%%%%%%%%%%%%%%%%%%%%%%%%%%%%
%\begin{tocentry}

%\centering
    
%\includegraphics[width=\textwidth]{TOCsmall.png}
   
% Inside the \texttt{tocentry} environment, the font used is Helvetica
% 8\,pt, as required by \emph{Journal of the American Chemical
% Society}.

% The surrounding frame is 9\,cm by 3.5\,cm, which is the maximum
% permitted for  \emph{Journal of the American Chemical Society}
% graphical table of content entries. The box will not resize if the
% content is too big: instead it will overflow the edge of the box.

% This box and the associated title will always be printed on a
% separate page at the end of the document.

%\end{tocentry}

%%%%%%%%%%%%%%%%%%%%%%%%%%%%%%%%%%%%%%%%%%%%%%%%%%%%%%%%%%%%%%%%%%%%%
%% The abstract environment will automatically gobble the contents
%% if an abstract is not used by the target journal.
%%%%%%%%%%%%%%%%%%%%%%%%%%%%%%%%%%%%%%%%%%%%%%%%%%%%%%%%%%%%%%%%%%%%%
\begin{abstract}

The structure of the excess proton in liquid water has been the subject of
lively debate from both experimental and theoretical fronts for the last
century. Fluctuations of the proton are typically interpreted 
in terms of limiting states referred to as the Eigen and Zundel 
species. Here we put these ideas under the microscope taking advantage
of recent advances in unsupervised learning that use local atomic descriptors to characterize environments of acidic
water combined with advanced clustering techniques. Our agnostic
approach leads to the observation of only a single charged cluster and 
two neutral ones. We demonstrate that the charged cluster involving the
excess proton, is best seen as an ionic topological defect in water's 
hydrogen bond network forming a single local minimum on the global free-energy
landscape. This charged defect is a highly fluxional moiety where the
idealized Eigen and Zundel species are neither limiting configurations
nor distinct thermodynamic states. Instead, the ionic defect enhances the presence 
of neutral water defects through strong interactions with the network.
We dub the combination of the charged and neutral defect clusters as 
\emph{ZundEig} demonstrating that the fluctuations between these local
environments provide a general framework for rationalizing more descriptive 
notions of the proton in the existing literature. 
%as either a distorted-Eigen cation or a more Zundel-like moiety.

\end{abstract}

%%%%%%%%%%%%%%%%%%%%%%%%%%%%%%%%%%%%%%%%%%%%%%%%%%%%%%%%%%%%%%%%%%%%%
%% Start the main part of the manuscript here.
%%%%%%%%%%%%%%%%%%%%%%%%%%%%%%%%%%%%%%%%%%%%%%%%%%%%%%%%%%%%%%%%%%%%%

%\begin{itemize}
%    \item The structure of the proton in liquid water has been a subject of a lot of debate over the last century: Eigen vs Zundel (localized vs shared proton)
%    \item We decided to take an agnostic approach to automatically discover what the proton structure in water is using an unsupervised learning protocol.
%    \item We use existing trajectories in the literature of the excess proton: 1) in water, 2) HCL concentrated to do our analysis.
%    \item 
    
%\end{itemize}

\section{Introduction}

Proton transfer (PT) reactions are at the heart of acid-base chemistry processes relevant for a wide range of phenomena in physical, chemical and biological systems~\cite{kreuer1996proton,cukierman2006et,wraight2006chance,raymond2011,Yamaguchi2014,HBN_watersplitting,gomes2014,Warburton2021,Creazzo2023,ChemRev3,stuchebrukov,halle_pnas2015,livoth2021}. The
thermodynamics associated with PT is central to determining pH, the protonation states of organic molecules in solution. The mechanisms by which protons move through hydrogen-bonded networks (HBNs) such as water, has attracted the attention of experimentalists and theoreticians alike~\cite{Hynes1997,ChemRev3}. Unlike other ions which diffuse in water via hydrodynamic diffusion, protons migrate within the HBN through the interconversion of covalent and hydrogen (H-)bonds, commonly referred to as the Grotthuss mechanism~\cite{Grotthuss1806,Agmon1995}. 

Textbook chemistry teaches us that the proton in liquid water does not exist as an isolated
entity as it does in the gas phase, but instead, electronically interacts with water~\cite{Atkins}. Since the mid 20th century, the structure of the proton in liquid water has been discussed in terms of two apparently limiting structures, namely, Eigen and Zundel~\cite{eigen1958self,zundel1969hydration}.
Manfred Eigen proposed that the excess proton is delocalized on a single water molecule forming the
(H$_3$O)$^+$ moiety. This species is bound via tight H-bonds to three surrounding water molecules, resulting in the (H$_{5}$O$_{9}$)$^+$ complex~\cite{eigen1964proton}.
On the other hand, George Zundel presented a more delocalized form of the proton suggesting, instead, that it is more evenly-shared between two flanking water molecules forming the (H$_{5}$O$_{2}$)$^+$ structure~\cite{zundel1969hydration}. 
This feature was invoked to rationalize the broad absorption in the IR region between $2000-3000$~cm$^{-1}$, commonly
referred to as the Zundel continuum~\cite{zundel1981,Zundel2000}. Since the mid-20th century these two assignments
of the proton structure have fuelled intense and lively debates regarding their
relative importance in the Grotthuss mechanism~\cite{botti2006,bakkerbonn2009,tokmakoff2015,dahms2016,kozari2021}. 

In a seminal paper in 1995, Agmon showed with very elegant back-of-the-envelope type calculations, that proton diffusion in water involves a rather low-activation barrier of approximately $2-3$~kcal/mol and that this is essentially coupled to changes in the local solvation of the excess proton\cite{Agmon1995} challenging previous ideas of a more collective phenomenon\cite{huckel1928}. A lot of our modern understanding of the proton structure in water has emerged from several decades of atomistic molecular dynamics simulations using both first-principles Density Functional Theory (DFT)~\cite{TuckermanMarxKleinParrinello1997,MarxTuckermanHutterParrinello1999,marx2010aqueous,tse_analysis_2015,HassanaliGibertiCunyKuhneParrinello2013,HassanaliGibertiSossoParrinello2014,cassone2020}, as well as reactive potentials such as empirical valence bond~\cite{LobaughVoth1996,PaveseVoth1998,LapidAgmonPetersenVoth2005,WuChenWangPaesaniVoth2008,KnightVoth2012,TseLindbergKumarVoth2015}. Quantum chemistry-type calculations in the gas phase and in cluster environments, which typically ignore thermal and entropic effects, have also played an important role in understanding the structure of the proton and how it is affected by hydration water~\cite{headrick2005spectral,yu2007vibrational,park2007eigen,DongNesbitt2006,SobolweskiDomcke2002,fournier2015snapshots,fournier2014,jordan2004}. 

%PT, its spectroscopic signatures and underlying mechanisms stretch far beyond just bulk water. Specifically, it is essential to understand how, at heterogeneous catalytic surfaces, water splitting affording oxygen evolution reactions takes place~\cite{raymond2011,Yamaguchi2014,HBN_watersplitting,gomes2014,Warburton2021,Creazzo2023}. Also in proteins, and more generally in biological systems, the thermodynamics and kinetics associated with PT play a critical role~\cite{ChemRev3,stuchebrukov,halle_pnas2015,livoth2021}. Furthermore, although it has recently been shown that the water self-dissociation is insensitive to nanoscale environments~\cite{solana23}, it is known that confinement suppresses the large mobility protons typically exhibit in the bulk~\cite{bakker2020}, a circumstance indicating the spatial extent of the fluctuations associated with proton migration. Thus, identifying the details underlying the structure and the behavior of the proton in water is important for understanding a wide class of different systems.

Different conclusions have been reached regarding the dominant structure of the proton including, besides the canonical Eigen and Zundel complexes, descriptions such as the \emph{distorted Eigen cation}~\cite{vothcalio2021} and the \emph{distorted Zundel structure}~\cite{fournierbroadband_2018,william2018,carpenter_entropic_2019}. These so-called limiting states also form part of the inherent language used to rationalize the Grotthuss mechanism where the proton diffusion is described in terms of the interconversion between Eigen and Zundel\cite{vuilleumier1998quantum,vuilleumier1999extended}. Since the identity of the proton changes as it migrates through the HBN, various types of chemically-inspired criteria need to be used to identify where the proton is instantaneously localized\cite{HassanaliGibertiCunyKuhneParrinello2013,HassanaliGibertiSossoParrinello2014,LapidAgmonPetersenVoth2005}. Within this context, chemical bias requires some form of dimensionality reduction which can lead to an uncontrolled information loss when examining fluctuations of the system in a reduced space of coordinates. In the last decade there has been a tremendous spurt in the use of data-science techniques to study complex molecular systems~\cite{glielmo2021unsupervised}. These approaches offer a framework in which patterns in the data emerging from the atomistic models are used to derive structural hierarchies and subsequently inform on the underlying physics and chemistry~\cite{glielmo2022ranking}. 

%  In this regard, the dominant proton transfer mechanism in bulk water is believed to be the Eigen-Zundel-Eigen (EZE) where proton diffusion occurs from the conversion of one Eigen species to another through a Zundel intermediate complex\cite{vothcalio2021}. Earlier studies suggesting a Zundel-Zundel mechanism were apparently exaggerated by the underlying model used\cite{vuilleumier1998quantum,vuilleumier1999extended}. In other contexts such as for example, water ionization, H-bonded cyclic water patterns mediating the interconversion of Eigen-like to Zundel-like cations have recently been proposed to emerge soon after autoprotolysis events from neural network potential simulations~\cite{huo2023}. Nonetheless, as Voth and co-workers have recently shown, Zundel-like geometries are not uncommon in distorted Eigen structures\cite{vothcalio2021} thus calling into question the sharp distinction between the two species. 

We have recently employed a battery of state-of-the-art data-driven techniques to study the nature of the fluctuations in liquid water at room temperature~\cite{offei2022high}. Therein, using the smooth overlap of atomic descriptors
(SOAP) to characterize local environments in liquid water and subsequently dimensionality
reduction and high-dimensional clustering, we showed that liquid water at room temperature is characterized by a single free-energy minimum that is effectively flat and broad with no evidence supporting the 2-state models of liquid water. In the current work, we extend this protocol to the study of the free-energy landscape of acidic water using previous \emph{ab initio} molecular dynamics (AIMD) simulations of HCl at various concentrations reported by Markland and co-workers~\cite{napoli2018}. The unsupervised clustering of the high dimensional SOAP space of HCl yields three statistically dominant clusters of which, only one corresponds to the excess proton. This charged cluster contains both Eigen and Zundel moieties, where these idealized structures do not emerge as limiting thermodynamic states. We demonstrate that the excess proton is best seen as a charged topological defect that enhances the concentration of neutral defects close to it which together can be chemically interpreted as a mix of the Eigen and Zundel species, which we refer to as \emph{ZundEig}. 

Section 1 begins with a summary of the computational methods employed in this work while Section 2 we describes the results that emerge from our unsupervised learning and the chemical interpretation given to the thermodynamic landscape. Finally we end in Section 3 with conclusions and future perspectives of our work.

\section{Methods}

%\begin{itemize}

%\item SOAP. General idea and equations. Similarity measure. Different references used. Parameters.
%\item ID
%\item DPA
%\item UMAP
%\item Classical and path integral AIMD trajectories of HCl 2M reported by Markland. (?)
%\end{itemize}

\begin{figure}[h]
\centering
\includegraphics[width=0.7\textwidth]{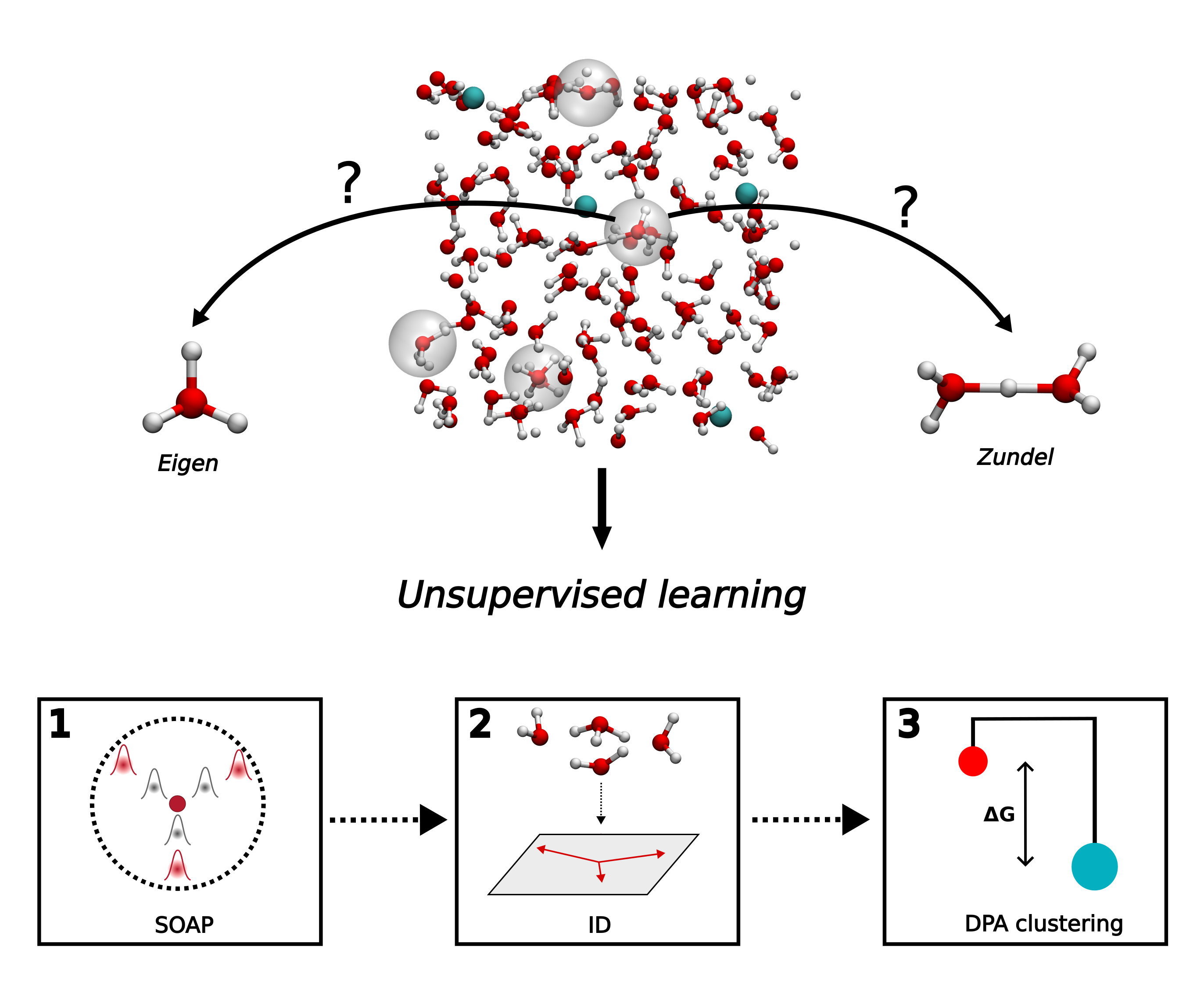}
\caption{Schematic protocol for the unsupervised learning that we perform on protonated water with the aim of determining whether the dominant species is an Eigen or Zundel. Trajectories obtained from AIMD simulations of protonated water are used to build local atomic descriptors (SOAP) (1) from which we extract the intrinsic dimension (ID) of the local solvation environment (2). This then serves as input for performing high dimensional clustering and extraction of the free energy (3).}
\label{scheme}
\end{figure}

Our protocol for performing the unsupervised learning of the molecular environments of protonated water follows closely our recent work on studying the free-energy landscape of liquid water at room temperature~\cite{offei2022high}. We take advantage of previously generated AIMD simulations by Markland and co-workers~\cite{napoli2018}. The general framework for performing our unsupervised learning approach is summarized in Figure~\ref{scheme} which involves several steps. Firstly, we build up the SOAP descriptors to characterize local environments~\cite{bartok2013representing}. Subsequently, the SOAP descriptors serve as input for computing the intrinsic dimension (ID) using the Two-NN method~\cite{facco2017estimating}. Finally, this ID is used to perform high dimensional free-energy construction~\cite{rodriguez2014clustering,rodriguez2018computing,d2018automatic}. A complementary dimensionality reduction technique using UMAP\cite{mcinnes2018umap} provides a useful visualization of the results obtained with this framework.

\subsection{\emph{Ab Initio} Molecular Dynamics}

In this work, we analyzed AIMD trajectories of HCl solutions that were previously generated by Markland and co-workers~\cite{napoli2018}. These simulations were performed using the revPBE exchange-correlation DFT functional~\cite{PBE,revPBE} with D3 dispersion corrections~\cite{grimme2010consistent}. This set of simulations includes a single excess proton in bulk water as well as two other concentrations, namely 2~M and 4~M HCl. The box size for the 2~M and 4~M solutions was $14.905$~{\AA}, containing $106$ H$_2$O molecules and either 4 or 8 excess protons and Cl$^-$ anions, respectively. The trajectory with only a single excess proton was performed in a simulation box with side length equal to $12.42$~{\AA}, containing 64 water molecules. %These conditions represent the experimental densities of these solutions at 300 K. 
Unless otherwise stated, we focus most of our analyses throughout the manuscript on the trajectories stemming from the 2~M HCl aqueous solution as these are the ones where we had access to significantly more statistics. 

%As for the 2~M HCl solution, we also analyzed path integral AIMD simulations from Markland and co-workers~\cite{napoli2018}, which include nuclear quantum effects. These simulations were performed by employing 32 beads using a ring polymer contraction framework~\cite{revPBE0}. The same simulation conditions as for AIMD with classical nuclei were used, with exception of the density functional (i.e., the hybrid revPBE0-D3~\cite{revPBE0,revPBE0D} instead of the revPBE-D3), as detailed in Ref.~\cite{napoli2018}. 

\subsection{Smooth Overlap Atomic Positions (SOAP)}

Over the last decade, SOAP has emerged as a powerful technique for describing the local environments around atoms and molecules including organic and inorganic materials ~\cite{maksimov2021conformational,grant2020network,de2016comparing,appignanesi2009evidence}. In addition, it has also recently been applied to investigate the properties of liquid water~\cite{offei2022high,capelli2022ephemeral,monserrat2020liquid}. The SOAP starting point is that of treating the local density of an atomic environment $\chi$ as a sum of Gaussian functions with variance $\sigma^2$, centered on all species that are neighbors of the central atom:
\begin{equation}
    \rho_{\chi} (\vec{r}) = \sum_{i \in \chi} \exp{\left(   \frac{-|\vec{r_i} - \vec{r}|^2}{2\sigma^2}\right)} \, \, \, .
    \label{dens_ll}
\end{equation}
The atomic density in equation~(\ref{dens_ll}) is then expanded in terms of spherical harmonics and radial basis functions such that:
\begin{equation}
    \rho_{\chi}(\vec{r}) = \sum_{n = 0}^{nmax} \sum_{l = 0}^{lmax} \sum_{m = -l}^{l} c_{nlm} g_n (r) Y_{lm} (\theta, \phi) \, \, \, .
\end{equation}
%where $c_{nlm}^{Z_i}$ are the coefficients of the expansion.
Upon accumulating the expansion coefficients, a rotationally invariant power spectrum is constructed:
\begin{equation}
\label{ps}
    p_{nn'l} (\chi) = \pi \sqrt{\frac{8}{2l+1}}\sum_m (c_{nlm})^{\dagger} c_{n^{'}lm} \, \, \, .
\end{equation}
Equation~(\ref{ps}) defines the components of the SOAP features that encode the information of the atomic environments that arise in the simulations of the excess proton.

In this work, to build the SOAP descriptors, we used the DScribe package~\cite{himanen2020dscribe}. A total of 265000 SOAP descriptors were constructed from the 2M HCl simulations. In order to employ the SOAP descriptors several hyperparameters must be set: 1) The species whose density we would like to describe, 2) the cutoff radius ($r_{cut}$) that defines the region that is considered as the environment $\chi$ of an atom, 3) the $\sigma$ value entering into Equation \ref{dens_ll} ,and 4) the number of radial and angular basis functions employed on the description of the densities ($n_{max}$, $l_{max}$). 

%In order to allow for a direct comparison of the analysis from the classical nuclei trajectories to that generated by means of the path-integral formalism, SOAP descriptors were placed on the centroid only. A total of 14000 descriptors were built for the path-integral runs.

%Multiple frames were taken from the trajectory every 200 fs to minimize the temporal correlations between different structures but at the same time providing sufficient data points in order to perform relevant statistical analysis. 

%By using equation \ref{deigen} it is possible to measure the similarity of one atomic environment to a reference one. In this sense, we constructed 3 different sets of SOAP descriptors using different parameters (Table \ref{param}) in order to compare the configurations extracted from the dynamics to one of 3 reference environments ($\chi'$):

%For practical purposes, one has to define 

Thus, to build the SOAP environments, we used the oxygen atoms as centers including both oxygen and hydrogen species within a radial cutoff of $3.7$~{\AA}. The $\sigma$ value chosen for the Gaussian function was $0.25$~{\AA}, consistent with a recent study of our group~\cite{laiodonkorhassanali2023} that showed the importance of using a smaller value to characterize local water structure. This ensures that the SOAP descriptor contains important information about the HBN structure. Finally, $n_{max}=12$ and $l_{max}=8$ were set to build the descriptors.

%\renewcommand{\arraystretch}{1.5}
%\begin{table}
%\caption{Parameters used to build the SOAP power spectrum for different %references.}
%\label{param}
%\begin{tabular}{ c | c | c | c }
%\hline
%  \textbf{Parameter} & \textbf{Solvated Eigen} & \textbf{Eigen} & \textbf{Zundel} \\
%\hline
%  center & O & O & H \\
%  $\sigma$ & 0.25 \AA{} & 0.25 \AA{} & 0.25\AA{} \\
%  $r_{cut}$ & 3.7 \AA{} & 2.15 \AA{} & 2.8 \AA{} \\
%  nmax & 12 & 10 & 10 \\
%  lmax & 8 & 6 & 6 \\
%\hline
%\end{tabular}
%\end{table}

\subsubsection{Intrinsic Dimension (ID)}

%Many times high dimensional data can be embedded in a lower dimensional space with minimal information loss. This is usually the case for data describing molecular systems where there are many constraints and correlations (bonds, angles, etc.) between different features. The Intrinsic Dimension represents the minimal set of features that can describe the data without significant information loss. Correctly estimating the ID allows to perform different types of analysis on the data in this space rather than in the full embedding space, which simplifies the calculations.

High-dimensional data generated from molecular systems are often characterized by correlations, implying that the system of interest resides in a lower dimensional manifold~\cite{glielmo2021unsupervised}. In the context of our analysis here, the power spectra associated with the SOAP descriptors are made up of high dimensional vectors consisting of over 1000s of elements. In previous studies, we have shown that this corresponds to the number of independent directions needed to characterize the fluctuations of the local atomic environments of water\cite{offei2022high} as well as in proteins\cite{sormani2019explicit}.

In this work, we determined the ID using the Two-NN estimator\cite{facco2017estimating} which uses information on the first and second neighbors of each data point. This method is based on the assumption of local uniformity of the data set. It can be demonstrated that, if this condition is satisfied, the ratio of the first and second neighbor distances ($r_1$ and $r_2$, respectively) is characterized by the following distribution:
\begin{equation}
    P(\mu) = \frac{d}{\mu^{d+1}} \, \, \, ,
\end{equation}
with $\mu$=$r_1$/$r_2$ and $d$ being the ID. If the sampled $\mu_i$s are independent, the ID can be calculated by Equation~(\ref{id}).
\begin{equation}
\label{id}
    d = \frac{N}{\sum_i\log(\mu_i)}
\end{equation}

\subsubsection{DPA clustering}

The calculation of the free-energy surface (FES) of molecular systems involves knowing the relevant coordinates capable of representing the different (meta)stable states of the system. Usually, the FES is estimated using a reduced number of coordinates assuming that these latter will suffice to capture the relevant fluctuations of the system one wants to represent. However, this dimensionality reduction unavoidably leads to information loss. In this work, we use the Point Adaptive k-Nearest (PAk) estimator that has been shown to correctly recover the FES of different molecular systems\cite{DERRICO2021476}. This method is an extension of the standard k-nearest estimator, in which the local density of every point $i$ in the data set is calculated as:
\begin{equation}
    \rho_i = \frac{\hat{k_i}}{r_{i,k}^d} \, \, \, ,
\end{equation}

where $d$ is the ID of the data set extracted from the Two-NN estimator and the $\hat{k_i}$s are chosen to be maximal under the constraint that the density within $r_{i,k}$ can be considered constant. As shown in Ref.\cite{DERRICO2021476}, this selection provides a good bias-variance trade-off in the density estimation. The free energies are then calculated as $F_i=-log(\rho_i)$. Once the point-dependent free energies $F_i$ have been determined, we then employ the Density Peaks Advanced (DPA) algorithm to find the minima on the FES, often referred to as \textit{clusters}. This method automatically determines the clusters and identifies only those ones that are within a statistical confidence interval $Z$, which is an adjustable parameter. This parameter determines when a density maximum can be considered a consequence of a fluctuation due to limited sampling or a real cluster. In practice, we set the Z parameter as the minimum value that results in a consistency between the results on two independently generated data sets.

Finally, in order to visualize the features of the data set as well as the results of the PAk-DPA algorithms, we used the Uniform Manifold Approximation and Projection (UMAP)\cite{mcinnes2018umap,diaz2019umap}. The UMAP method gives a low-dimensional projection of a high-dimensional data set (similar to t-SNE~\cite{tsne}) preserving its topography as much as possible and therefore allowing the visualization of the main features of the high-dimensional FES.

\section{Results}

\subsection{Free-Energy Landscape of 2M HCl}

We determined the SOAP descriptors centered on all the oxygen atoms with a cutoff of $3.7$~{\AA} including both oxygens and hydrogens of the environment. Such a choice ensures that the first solvation shell of each putative molecule or ion in the system is included in the atomic environment. We subsequently computed the ID of the SOAP descriptors, obtaining a value of $\sim$33. With these two inputs (i.e., the SOAP descriptors and its ID) we then performed the Density Peaks Advanced (DPA) clustering with a confidence interval of $Z=3.0$ obtaining two major clusters, one enclosing 91\% of all data points and the other 8\% of all the data points. The remaining 1\% of the data does not belong to any of the two clusters. The presence of two clusters is consistent with our intuitive expectation that in water with an excess proton there are likely to be at least two broadly speaking, chemical topologies: one corresponding to neutral water molecules and another to the excess proton chemically bound in some fashion to a neutral water molecule. 

\begin{figure}[h]
\centering
\includegraphics[width=0.8\textwidth]{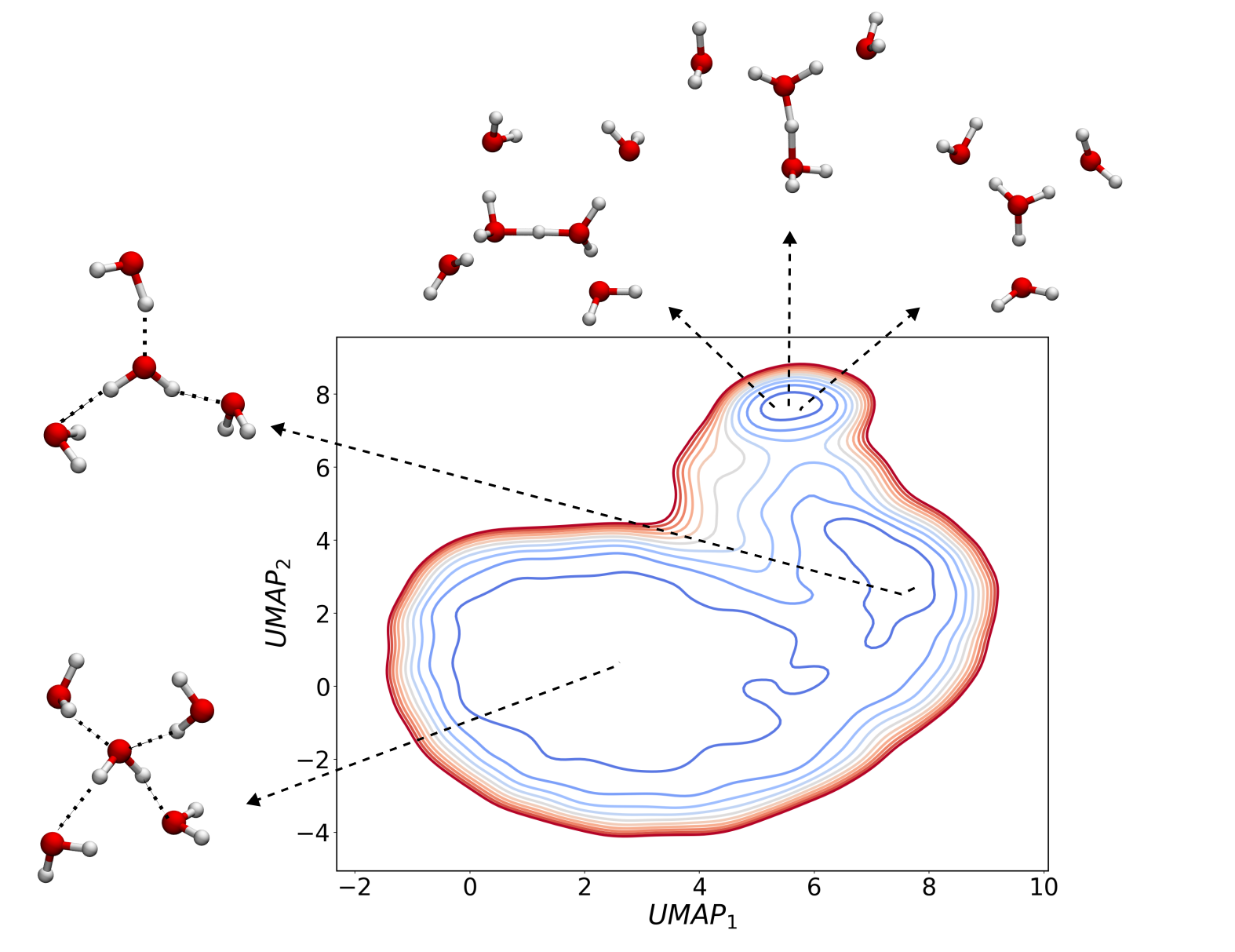}
\caption{Two-dimensional contour plot obtained from the projection of the SOAP space onto two UMAP variables. Three distinct minima are observed and some of the representative molecular structures associated to them are reported.}
\label{fes}
\end{figure}

%which included over the $99\%$ of all data points. This suggests that in the sample there are two statistically significant structures associated with the oxygen atoms and their own local environments.
In order to build our intuitions and visualize the underlying topography high-dimensional free-energy landscape we projected the SOAP descriptors using UMAP.  Figure~\ref{fes} shows a contour plot of the resulting 2D-projection where we observe the presence of three distinct minima. Each of these minima essentially corresponds to different molecular patterns present in the system. Upon visual inspection of the atomic environments associated with each minimum, one of these patterns corresponds to neutral -- on average tetrahedrally coordinated -- water molecules (bottom left), another one to charged water environments including the excess proton (top) and, finally, the third minimum involves neutral defect-looking water molecules (middle right). 

\begin{figure}[h]
\centering
\includegraphics[width=1.0\textwidth]{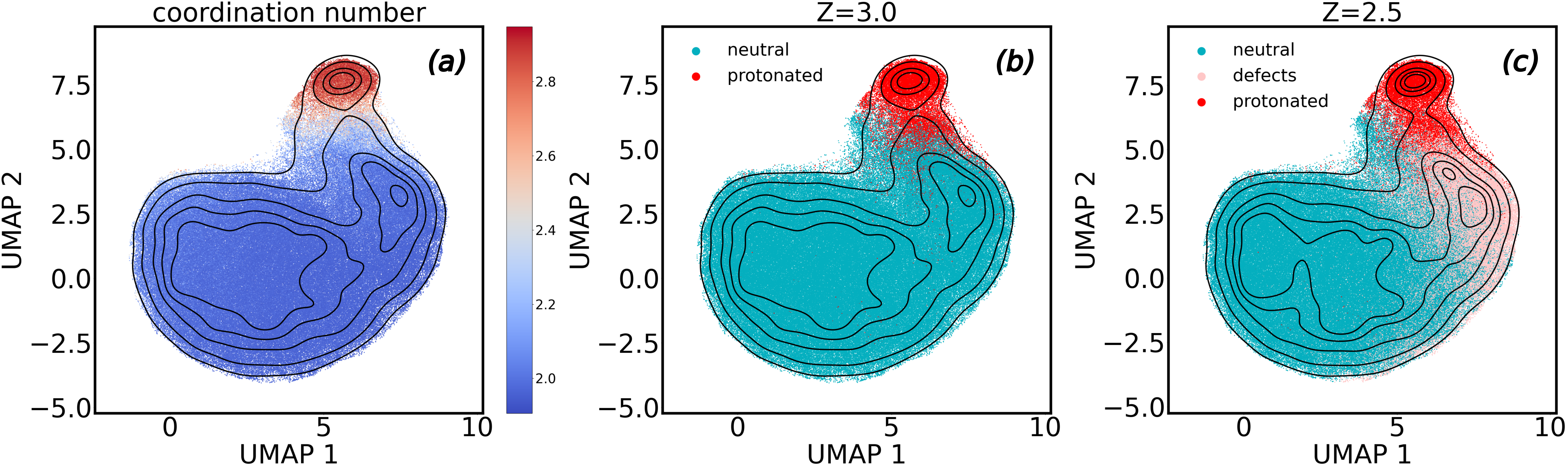}
\caption{Two-dimensional UMAP projections coloring data by coordination number (a) and by cluster labels obtained with DPA clustering analysis (b,c). Two confidence levels are displayed: $Z=3.0$ (b) and $Z=2.5$ (c).}
\label{clustering}
\end{figure}

To better quantify the structural origins of the different minima, we constructed a standard coordination number using a switching function defined as $n_{H} = \sum_{i=1}^{N_{H}} \frac{1}{\exp{\left[\kappa(r-r_c)\right]+1}}$, which counts the number of hydrogen atoms around every oxygen atom within a given radial cutoff. This allows for a quick identification of the oxygen atoms having either two or three covalently-bonded hydrogens, allowing for an initial assignment of the clusters. Note that $N_H$ is the total number of H atoms in the system whereas $\kappa$ and $r_c$ are the parameters of the smooth step function taken from the literature to describe the hydrogen coordination of a central oxygen atom~\cite{SPRIK2000139}. Figure~\ref{clustering}a) shows the same UMAP projection, colored as a function of $n_{H}$. We observe a gradient in the coordination number starting from $\sim$3 and approaching $\sim$2 as one moves from the top to the bottom. This analysis confirms the results obtained from visual inspection of the trajectories which are shown in the snapshots of Figure~\ref{fes}. The minimum observed associated with higher coordination numbers ($\sim$3) corresponds to the typical molecular environments that would be associated with the excess proton whilst the other two minima -- which exhibit a lower coordination number ($\sim$2) -- are those associated with neutral water molecules.

Although the UMAP projection seems to suggest the presence of three distinct minima, our DPA analysis shows the existence of only two clusters using Z=3.0. The granularity of the underlying free-energy landscape -- and therefore the number of clusters -- depends on the choice of the confidence parameter Z. Although lowering the value of Z leads to the emergence of more clusters, the confidence of their statistical significance is reduced\cite{d2018automatic} and thus, has to be combined with a chemical and physical interpretation. To examine how our results depend on the Z parameter, we have compared the outcome of the clustering obtained for $Z=3.0$ and $Z=2.5$. We observe that with $Z=2.5$, we obtain three distinct clusters which include 67\%, 19\% and 8\% of all data points, respectively. Subsequently, we have used the same reduced representation of UMAP, coloring the data points with the labels obtained from the clustering, as shown in Figure~\ref{clustering}b) and c), which  compare the results for $Z=3.0$ (middle panel) to those for $Z=2.5$ (rightmost panel), respectively. 

Interestingly, for $Z=3.0$ we observe that one of the clusters obtained by means of DPA partially overlaps with the minimum involving a higher coordination number ($\sim3$) while the other cluster is dominated by a coordination number of 2. Based on these features, the two clusters can be assigned as the protonated and neutral water clusters, respectively. Figure~\ref{clustering}c), instead, shows that the effect of reducing the Z parameter and re-coloring the UMAP landscape with the labels of the three clusters is that of splitting the large neutral water cluster into two regions which essentially cover the two minima associated with neutral water. This third cluster, located in the middle of the UMAP projection, is characterized by a coordination number of $2$ indicating that it has a local topology of a neutral water molecule. 

With the aim of characterizing the origins of this cluster, we have studied the H-bond patterns associated with these environments. Interestingly, water molecules in the new third cluster exhibit a deficiency in the number of donating and accepting H-bonds, implying that they are neutral water coordination defects~\cite{GasparottoHassanaliCeriotti2016,henchman2010topological}. Specifically, relative to neutral water, these defects have a reduced tendency to accept H-bonds (see SI Figure S1). In order to understand whether these defect water molecules have a tendency to form in close proximity to the excess proton, we determined the pair-correlation function between the oxygen atoms centers of the protonated and the two neutral water clusters (see SI Figure S2). We find that these neutral defective water molecules prefer to lie closer to the excess proton compared to waters coming from the other neutral basin. This includes water molecules located both at H-bonding distances as well as at length-scales consistent with the interstitial water molecules, the latter of which originates from neutral water defects of neighboring protons (see SI Figure S2). 

%The physical rationale underlying this observation lies in the fact that the excess proton increases the population of defect-like water molecules, which are more prone to accept the proton via the Grotthuss mechanism~\cite{Agmon1995,LapidAgmonPetersenVoth2005}.

In a nutshell, these results demonstrate that the unsupervised approach going from the SOAP descriptors to the free-energy clustering allows for an automatic identification of local environments in protonated liquid water, a feature that is consistently reproduced in path-integral simulations, different concentrations of the HCl, and also with more accurate treatment of the electronic structure (see SI Figure S3).  The existence of the presence of a defect neutral water pattern as a cluster forming close to the excess proton is fully consistent with the ``special-pair dance'' proposed by Voth and co-workers, where a strong interaction is formed between the excess proton and its nearest neighbor water molecule~\cite{markovitch2008special}. Within our unsupervised framework can be rationalized as the highly fluxional protonated cluster interacting and creating the neutral-defect. However, we do not find any evidence for the existence of two different clusters that correspond to Eigen and Zundel species. Although reducing the confidence parameter to $Z=2.0$ leads to the formation of more clusters inside the protonated basin, these inner clusters do not segregate into any statistically significant minima resembling either the Eigen or the Zundel moiety (see SI Figure 4).

\subsection{Eigen or Zundel?}

The extent to which the proton delocalizes in the H-bond network is sensitive to factors such as temperature, zero-point energy fluctuations and hydration. At low temperatures and in isolated (gas-phase) clusters, the idealized Eigen and Zundel geometries have been proven to be more easily distinguishable~\cite{jordan2004,fournier2014,KuligAgmon2014,fournier2015snapshots}. Agmon and co-workers have shown that in the gas phase, a proton added to the water tetramer yields the (H$_3$O)$^+$ while for the water dimer, it yields (H$_{5}$O$_{2}$)$^+$\cite{}. Similarly, Johnson and co-workers have shown that the excess proton with 20 water molecules, often referred to as magic number protonated clusters, involves the Eigen cation ((H$_3$O)$^+$) on the surface~\cite{johnson2014}. In addition, some of the early path-integral molecular dynamics simulations by Parrinello and co-workers demonstrated that the inclusion of nuclear quantum effects resulted in more proton delocalization thereby resulting in more Zundel-like geometries\cite{TuckermanMarxKleinParrinello1997,MarxTuckermanHutterParrinello1999}. 

To provide a more agnostic characterization of the similarity of the condensed phase proton structures to Eigen or Zundel geometries, we identified relevant milestone structures sampled from simulations in the gas phase, where the structures can be identified as Eigen or Zundel without any ambiguity similar to the geometries illustrated in Figure 1.  For more details on how the reference structures are chosen and built, the reader is referred to the SI. In summary, this is a post-processing procedure where we recompute the SOAP descriptors using the Eigen and Zundel configurations obtained from 20 randomly chosen configurations in the magic number protonated simulations. Among these Eigen geometries, the average O-H bond length is 1.05~\AA{}, while in the Zundel, the flanking water molecules have an average O-H bond of 1.01~\AA{}. Furthermore, in these Zundel reference configurations, the proton transfer coordinate is chosen to be 0 with a threshold of 0.01~\AA{} corresponding to situations where the proton is shared equally. Using these new SOAP vectors, we have computed the similarity measure $d(\chi,\chi')$ between the protonated and the neutral clusters obtained with $Z=3.0$ ($\chi$) at 2~M HCl, with respect to these reference configurations ($\chi'$). $d(\chi,\chi')$ is chosen to be the minimum distance obtained from the 20 configurations. This yields two distances, namely $d_{eigen}$ and $d_{zundel}$. 

\begin{figure}[h]
\centering
\includegraphics[width=1.0\textwidth]{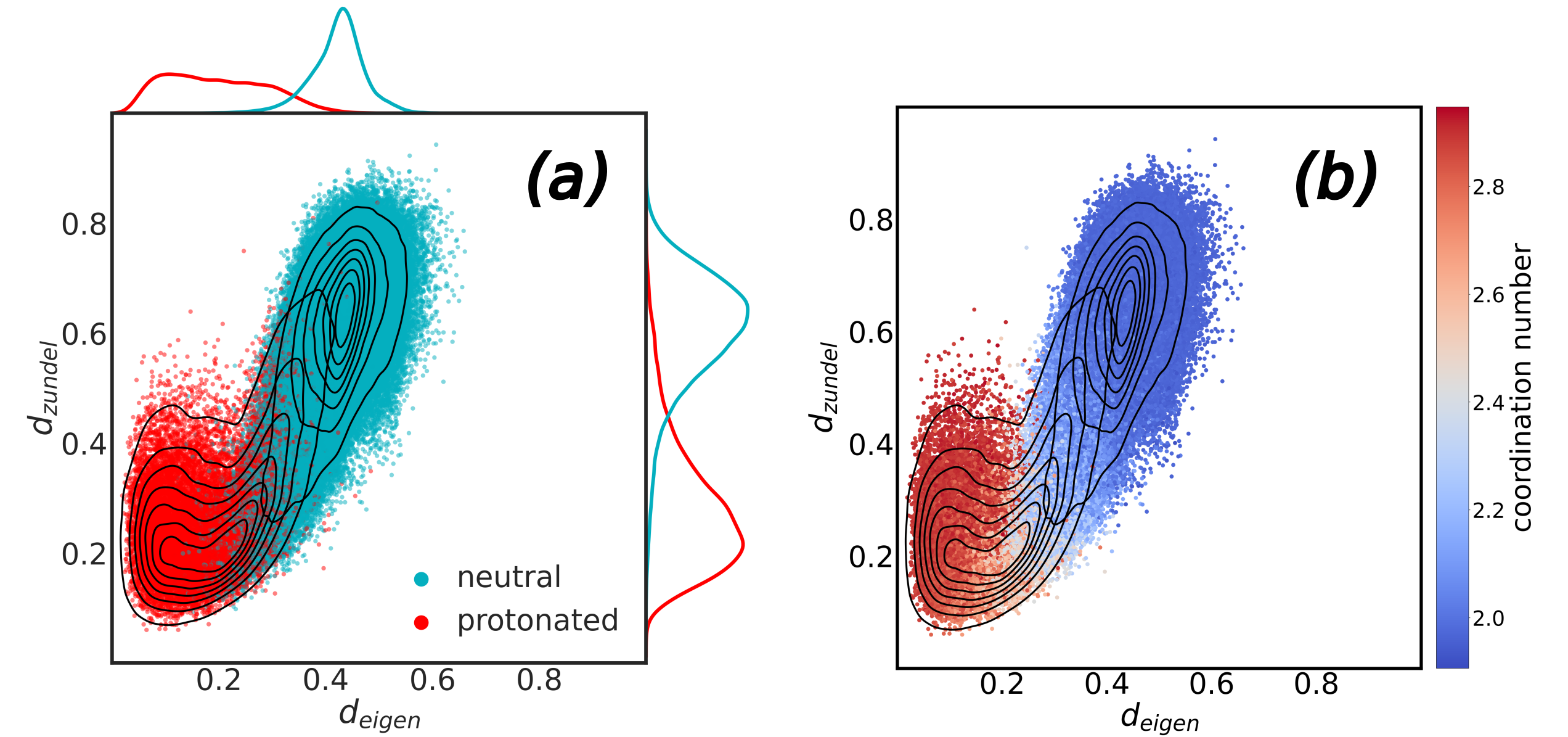}
\caption{2D density plot showing the SOAP distance to the idealized Eigen and Zundel references for the protonated and neutral clusters. The protonated cluster is characterized for smaller distances (structures are more similar to the references) than the neutral cluster, although a big overlap is observed between both of them. Coloring the distribution by coordination number shows that the protonated configurations have a coordination number $\sim3.0$ whereas the neutral ones $\sim2.0$.}
\label{deigen}
\end{figure}

Figure~\ref{deigen}a) shows the distribution of distances to the Eigen ($d_{eigen}$) and Zundel ($d_{zundel}$) reference structures both for the protonated and neutral clusters. A substantial overlap in the proximity of the protonated cluster to Eigen- and Zundel-looking geometries is observed. Figure~\ref{deigen}b) illustrates the distribution of $d_{eigen}$ and $d_{zundel}$ colored now as a function of the coordination number $n_H$. In this manner, we see clearly that neutral water molecules are dominated by a coordination number close to 2 while that of the protonated cluster involves an enhanced coordination number of 3.  A continuous transition between the neutral and the protonated environments following the $d_{eigen}$ and the $d_{zundel}$ coordinates is observed resulting in a large overlap between the neutral and protonated clusters.  This arises from the fact that since the excess proton is a highly fluxional topological ionic defect, it strongly distorts the HBN so that water molecules in close vicinity to the excess proton can be classified as either a charged or neutral cluster.

%By combining the SOAP-derived distances with respect to the milestone structures as well as the $n_H$, we observe that the local water environments with higher coordination number resemble both Eigen and Zundel species, making a sharp distinction between the two questionable (Figure~\ref{deigen}b). 

%As eluded to earlier, NQE's through zero-point energy fluctuations have been shown to enhance the delocalization of the proton along hydrogen bonds\cite{}. In this regard, one may wonder how much the similarity measures shown in Figure \ref{deigen} change upon the introduction of quantum fluctuations. In Figure XXX, we compare in the left and right panels the $d_{eigen}$ and $d_{zundel}$ distances for the classical and quantum simulations respectively, the latter of which is obtained from the path integral simulations reported earlier. Note that for the PIMD simulations, the SOAP descriptors and DPA analysis is performed using only the centroid of the path integral after which the $d_{eigen}$ and $d_{zundel}$ is constructed using all the 32 beads of the path integral.

%While this clearly reinforces the previous observations that the similarity of the excess proton environments in 2~M HCl aqueous solutions to the two idealized geometries overlap in distance, $d_{eigen}$ appears to be characterized by a peak located at smaller distance values with respect to its reference milestone compared to $d_{zundel}$, a circumstance suggesting a slight preference for Eigen-like geometries.  

\begin{figure}[h]
\centering
\includegraphics[width=\textwidth]{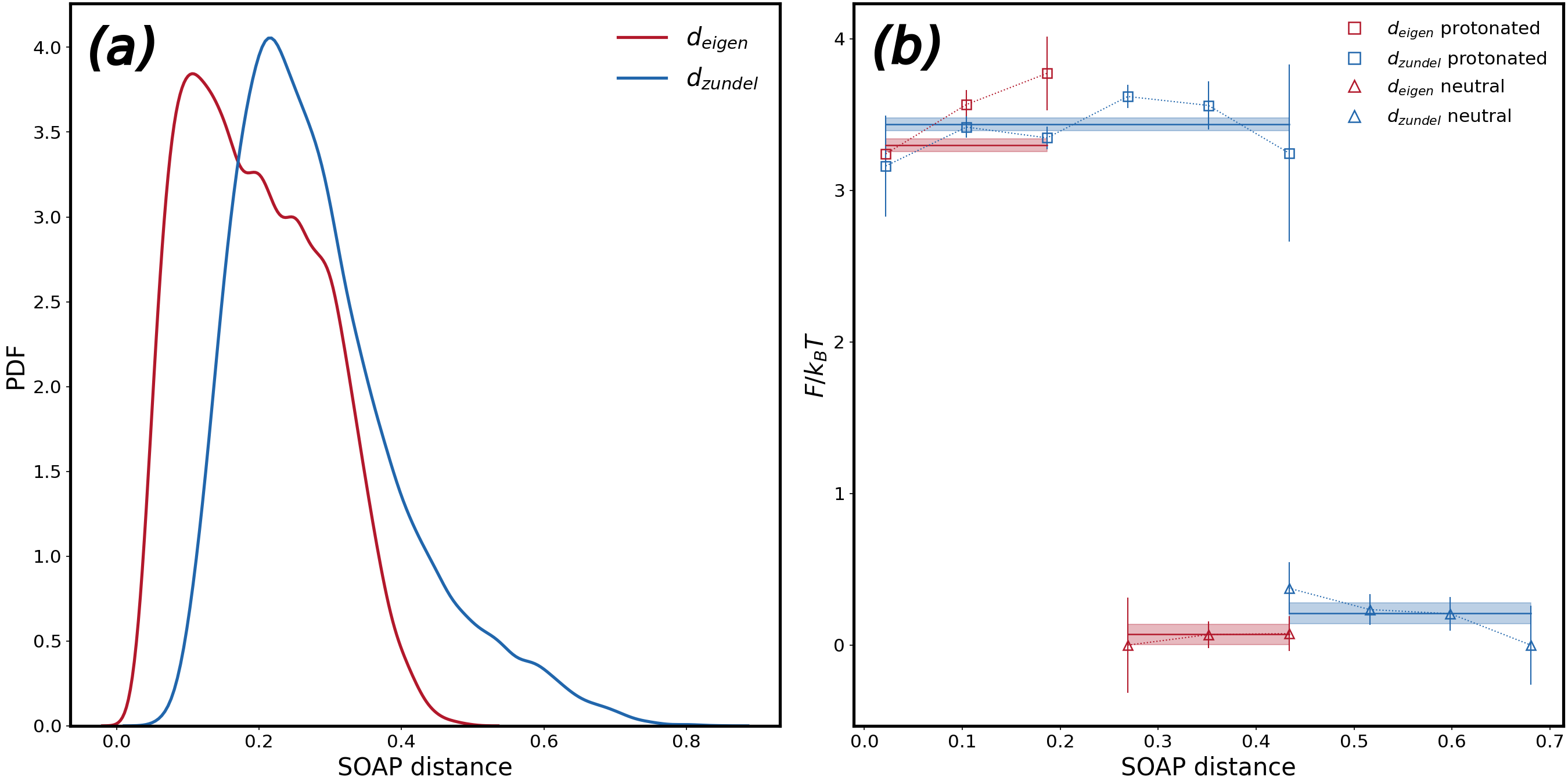}
\caption{(a) Density distributions for $d_{eigen}$ and $d_{zundel}$ calculated from all structures in the protonated cluster. (b) Free energies calculated in the SOAP space from the DPA clustering algorithm. Only the energies of the core points of the protonated (squares) and neutral (triangles) clusters are considered. The coordinates $d_{eigen}$ and $d_{zundel}$ were divided in bins and each point in the plot represents an average of free-energy values in each bin. Bars indicate the standard error of the sample in each bin. Horizontal lines follow the average free energy of the points in the protonated and in the neutral clusters as a function of $d_{eigen}$ and $d_{zundel}$. Shaded areas display the standard error.}
\label{densities}
\end{figure}

The scatter plots and contour lines of Figure~\ref{deigen}a) are slightly skewed in favor of Eigen-like configurations, as observed by the fact that the density peak in the protonated cluster of Figure~\ref{deigen}a) is closer to zero along the $d_{eigen}$ coordinate rather than along the $d_{zundel}$. This is illustrated more clearly in Figure~\ref{densities}a), which shows the one-dimensional distributions of the $d_{eigen}$ and $d_{zundel}$ distances.  The $d_{eigen}$ and $d_{zundel}$ distributions of Figure~\ref{densities}a) are one-dimensional projections from a high-dimensional distribution and, therefore, cannot be used to infer the dominance of Eigen vs Zundel. In order to understand the impact of this projection, we have determined the high-dimensional free energies as described earlier from $F_i=-log(\rho_i)$, of the protonated and neutral clusters as a function of $d_{eigen}$ and $d_{zundel}$. These free energies are illustrated in Figure~\ref{densities}b). In order to construct these curves, we have computed the average of the point free energies for core-points of each cluster along the $d_{eigen}$ and $d_{zundel}$ coordinates. The neutral environments lie in a free-energy basin located about $2-3~k_{B}$T lower than the protonated ones. Interestingly, this free-energy difference is fully consistent with the small activation energy required for proton migration~\cite{Agmon1995}. Examining the free-energy changes for the protonated cluster along $d_{eigen}$ and $d_{zundel}$ coordinates essentially confirms the findings obtained from the clustering analysis presented earlier. Within the statistical errors of our simulations, the free energy along these two variables is essentially flat. Furthermore, measuring the proximity of the excess proton using $d_{eigen}$ and $d_{zundel}$, it is clear that the free energies along these variables are also within statistical error of each other. This evidence implies that fluctuations of the excess proton in water occur on a \emph{flat} free-energy landscape, where transformations between Eigen- and Zundel-like configurations occur. Eigen- and Zundel-like geometries are however, neither limiting nor stable thermodynamics states on the free-energy landscape.

\begin{figure}[h]
\centering
\includegraphics[width=\textwidth]{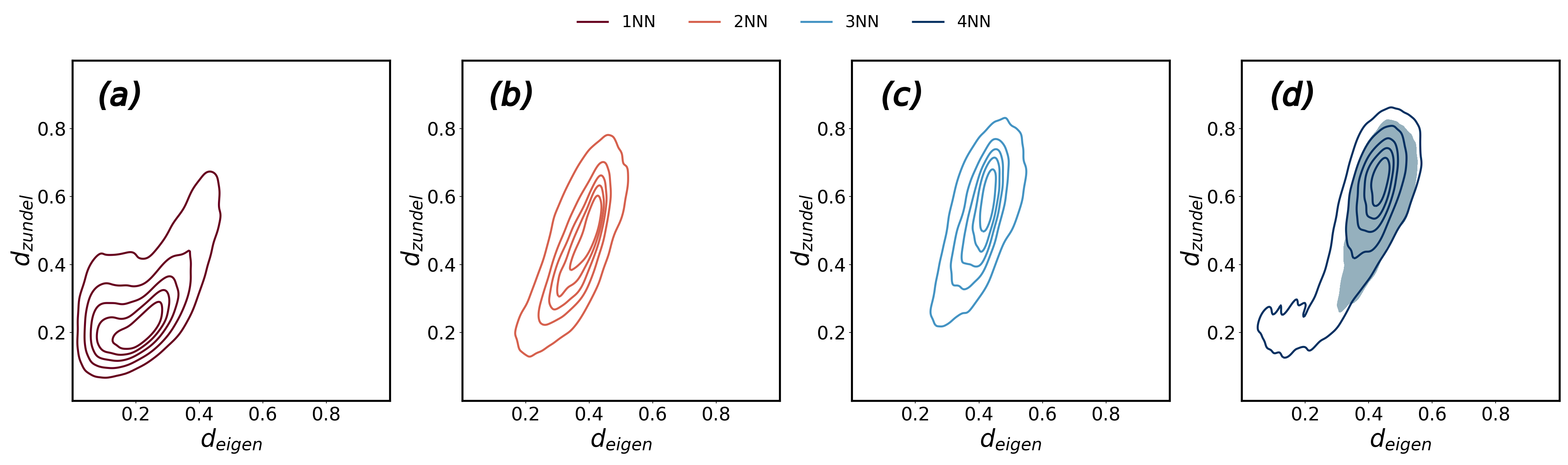}
\caption{$d_{eigen}$ (x-axes) and $d_{zundel}$ (y-axes) density maps referred to the first (a), second (b), third (c), and fourth (d) nearest neighbors of the molecules in the protonated cluster.}
\label{distNN}
\end{figure}

Up to this point, we have mainly focused on examining the structural similarities between the charged protonated cluster with respect to the Eigen and Zundel species. On the other hand, this ionic and highly fluxional topological defect strongly perturbs the nearby environment. It is hence interesting to investigate the extent of the similarity to the Eigen and Zundel species of the atomic environments in close proximity to the charged protonated cluster. To do so, we construct the $d_{eigen}$ and $d_{zundel}$ coordinates for the 4 nearest neighbors (NNs) of the charged cluster. For every snapshot of our trajectory we identify the 4 closest oxygen atoms to the charged protonated cluster, extracting the $d_{eigen}$ and $d_{zundel}$ coordinates for those selected oxygen centers. Figure~\ref{distNN} shows the density plots for $d_{eigen}$ (x-axes) and $d_{zundel}$ (y-axes) for the first, second, third and fourth NNs of the protonated cluster. The first NN (1NN), which essentially corresponds to the special-pair discussed in the literature~\cite{markovitch2008special}, exhibits a sizable overlap in the $d_{eigen}$ and $d_{zundel}$ distributions. This further corroborates the idea that within the special-pair, a sharp distinction between Eigen and Zundel molecular structures does not adequately describe the fluctuations of the proton. Furthermore, the fact that the 1NN displays substantial similarity to the Zundel is also fully consistent with vibrational spectroscopy measurements of HCl~\cite{fournierbroadband_2018,william2018,carpenter_entropic_2019}.  

Going beyond the 1NN, we find that for both 2NN and 3NN (Figure~\ref{distNN}b) and c)), the $d_{eigen}$ and $d_{zundel}$ distributions also exhibit a significant overlap, even though the sampled configurations appear to be slightly closer to the Eigen rather than to the Zundel complex. Thus, in addition to the special-pair atomic environment that resides close to the charged cluster (1NN), proton delocalization to other neighbors involves a substantial overlap between the Eigen and the Zundel configurations. The picture that emerges from the current analysis is in good agreement with previous characterizations of the PT mechanisms associated with the special-pair dance, where the excess proton dynamically fluctuates between forming H-bonds between three different atomic environments\cite{LapidAgmonPetersenVoth2005}. Finally, Figure~\ref{distNN}d) displays the distributions obtained for the 4NN, where the $d_{eigen}$ and $d_{zundel}$ coordinates essentially converge to those obtained for the neutral cluster (shown as shaded area).

\subsection{Combining SOAP and Chemical Descriptors}

\begin{figure}[h]
\centering
\includegraphics[width=\textwidth]{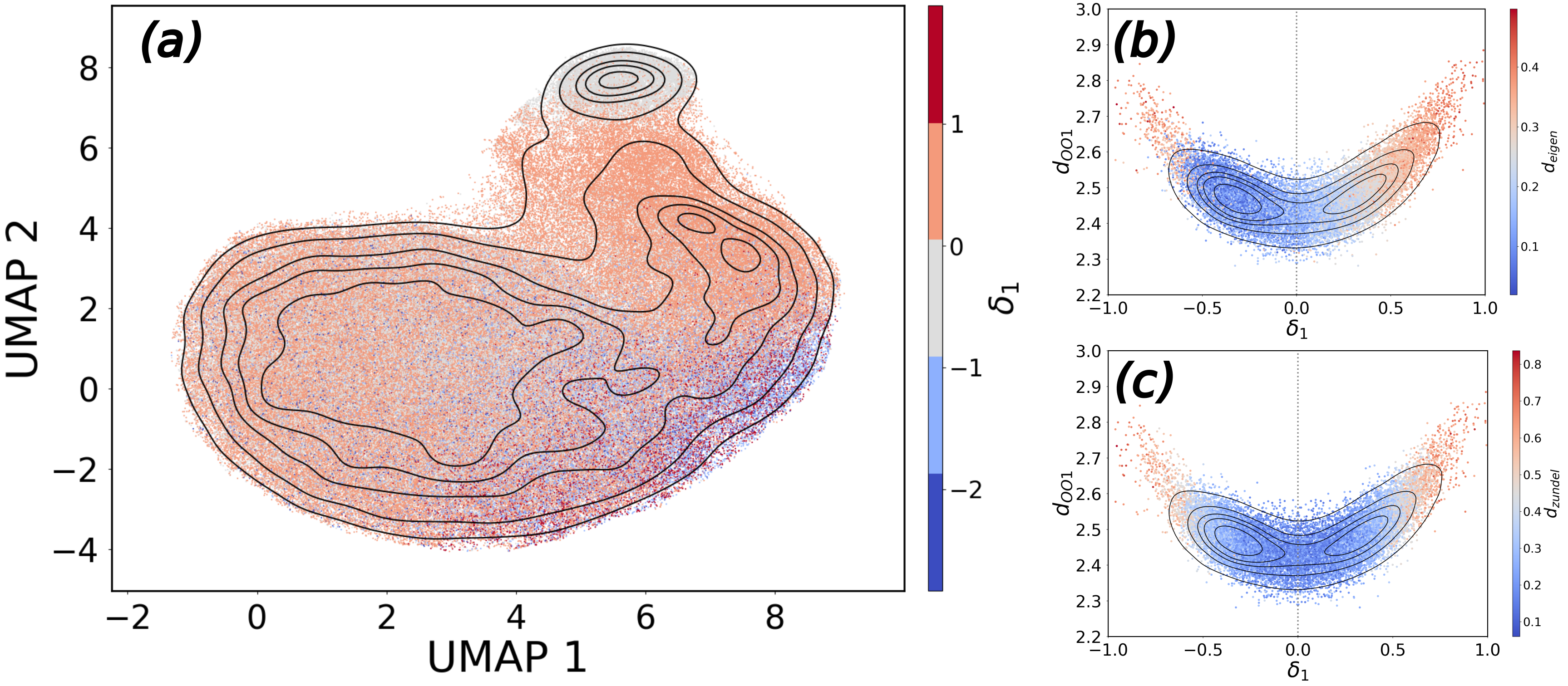}
\caption{UMAP projection of the SOAP vectors colored as a function of the proton sharing coordinate (PTC) $\delta_1$ (a). Distances of the first-neighbor oxygen atom from the protonated oxygen $d_{OO1}$ as a function of the PTC $\delta_1$ and colored by the SOAP distances $d_{eigen}$ (b) and $d_{zundel}$ (c).}
\label{struct}
\end{figure}

Due to the fact that the Grotthuss mechanism involves the interconversion of hydrogen and covalent bonds in a highly cooperative process~\cite{LapidAgmonPetersenVoth2005}, identifying the relevant degrees of freedom that describe the proton hopping phenomenon is challenging. The extent of PT along the H-bond is typically quantified through the proton transfer coordinate (PTC), which is measured by a difference in distance between the proton and the oxygen atoms along which the transfer occurs. In order to enhance the chemical interpretability of the SOAP projections in UMAP, we begin by showing in the left panel of Figure~\ref{struct} the original UMAP projection colored as a function of the PTC. To construct the PTC, we sit on the oxygen atom of each the three clusters identified with the confidence parameter $Z=2.5$, identify the 1NN and then compute the PTC as $d_{O_{C}H}-d_{HO_{1NN}}$, where O$_{C}$ is oxygen atom associated with the charged protonated cluster. In the bulk neutral water cluster, the PTC coordinate is randomly distributed, meaning that it takes on both positive and negative values which correspond to water molecules involved in accepting and donating H-bonds, respectively. 

Honing in on the charged protonated and neutral defect minima, the PTC is on average negative ($-0.4$) and positive ($0.5$), respectively. The interconversion between Eigen and Zundel species has typically been rationalized through the PTC: values close to zero are often interpreted as the creation of a Zundel cation, while deviations from this as the formation of the Eigen. We have shown instead (see Figure~\ref{deigen}) that the minima associated with the charged cluster involves a broad spectrum of configurations that can be identified in similarity to both the Eigen and Zundel ``states''. The neutral water defect structure forming the minima in the middle of the UMAP projection, appears to be dominated by PTC values that are on average greater than zero consistent with the notion of a molecular structure that is most likely to accept a proton through PT.

Besides the special-pair dance which emerges from our analysis, PT has also been shown to be strongly coupled to lower-frequency vibrational modes of the heavy atoms that sandwich the proton~\cite{MarxTuckermanHutterParrinello1999,HassanaliPrakashEshetParrinello2011,HassanaliGibertiCunyKuhneParrinello2013}. Figure~\ref{struct}b) and c) show the distributions of the first-neighbor oxygen atom distances from the protonated oxygen $d_{OO1}$ obtained along the PTC, which reproduce the banana-like profiles observed in many previous studies (see, e.g., Ref.~\cite{HassanaliGibertiCunyKuhneParrinello2013}). On each banana-shaped profile, we color the plot with $d_{eigen}$ (Figure~\ref{struct}b) and $d_{zundel}$ (Figure~\ref{struct}c) distances.  Interestingly, the asymmetric nature of the Eigen cation with respect to the location of the proton is consistent with its position on the banana-plot i.e, asymmetrically populated basins at short (blue dots) and long (orange dots) $d_{eigen}$ distances (Figure~\ref{struct}b). At the same time, however, PTC values close to zero can also resemble Eigen-like species. The banana-plot of the centrosymmetric Zundel complex as shown in Figure~\ref{struct}c), shows that there exists a very diffuse basin accessible to the excess proton across the full landscape. In fact, Figure~\ref{struct}c) clearly shows that the proton rattling phenomenon along the H-bonds is always featured by short $d_{zundel}$ coordinates. In other words, the Zundel cation remains a Zundel complex even when the PTC departs from zero, confirming once again that discretization of the excess proton into two species does not adequately account for the underlying relevant fluctuations. 

\section{Conclusions}

In the current work, we have exploited the smooth overlap of atomic descriptors (SOAP) along with advanced clustering techniques to characterize local environments in acidic water. Using previous AIMD simulations of HCl aqueous solutions at various concentrations reported by Markland and co-workers~\cite{napoli2018}, we show that the unsupervised clustering of the high-dimensional SOAP space yields three statistically dominant clusters of which, only one corresponds to the excess proton. This charged cluster contains both Eigen and Zundel configurations, although the idealized structures do not emerge as limiting thermodynamic states. By exploiting the definition of a metrics allowing for the evaluation of the distance $d$ from well-defined Eigen ($d_{eigen}$) and Zundel ($d_{zundel}$) gas-phase milestone structures, we also have determined the free-energy changes for the protonated cluster along the $d_{eigen}$ and $d_{zundel}$ coordinates. Interestingly, the free energy along these two variables is flat whilst the free-energy difference between the Eigen and the Zundel moieties is non-existent, implying that fluctuations of the excess proton in water take place on a flat free-energy landscape, where transformations between Eigen- and Zundel-like configurations easily occur. 

Another relevant feature emerging from our unsupervised analyses of the global free-energy surface of the system is that the protonated and neutral clusters are separated by a peculiar basin associated with non-tetrahedrally coordinated water molecules. In fact, the charged protonated cluster enhances the concentration of a specific type of neutral defect that tends to lie in close proximity to the proton. In particular, this neutral defect is characterized by a H-bond topology that is defective, a feature that is enhanced in HCl solutions compared to bulk water with excess protons.  Conversion between the charged and neutral cluster essentially describes the fluctuations of the proton involved in the Grotthuss mechanism where structural interconversions of a wide spectrum of molecular structures form part of a broad spectrum of fluctuations of a charged topological defect. Eigen and Zundel as originally conceived form just one of many different structures and perhaps it is time to let go of very specific assignments based on two-state type descriptions. 

Finally, when combining our clustering analysis with the SOAP distance metrics and previously used chemical-based coordinates, a clear scenario emerges. There exists a very diffuse basin accessible to the excess proton associated with a highly fluxional charged defect whose chemical identity is fuzzy. In other words, the proton rattling phenomenon along the H-bonds is always featured by moieties which are, simultaneously, Eigen- and Zundel-like, a circumstance which reinforces the notion that two-state cataloging does not provide a correct physical description of the system. Instead, fluctuations between protonic and neutral defects -- that we dub as \emph{ZundEig} -- is a more appropriate representation for studying PT mechanisms. We believe our results provide a general framework for examining the evolution of the structure of proton in more complex environments such as at interfaces\cite{chemrevinterfaces2016,creazzo2020}, under confinement\cite{solana23,bakker2009}, and also in the presence of external electric fields where the extent of proton delocalization can be modulated\cite{cassone2020,saitta2012,cassonejpcb14}.

\clearpage

\bibliography{bib}

\providecommand{\latin}[1]{#1}
\makeatletter
\providecommand{\doi}
  {\begingroup\let\do\@makeother\dospecials
  \catcode`\{=1 \catcode`\}=2 \doi@aux}
\providecommand{\doi@aux}[1]{\endgroup\texttt{#1}}
\makeatother
\providecommand*\mcitethebibliography{\thebibliography}
\csname @ifundefined\endcsname{endmcitethebibliography}
  {\let\endmcitethebibliography\endthebibliography}{}
\begin{mcitethebibliography}{95}
\providecommand*\natexlab[1]{#1}
\providecommand*\mciteSetBstSublistMode[1]{}
\providecommand*\mciteSetBstMaxWidthForm[2]{}
\providecommand*\mciteBstWouldAddEndPuncttrue
  {\def\EndOfBibitem{\unskip.}}
\providecommand*\mciteBstWouldAddEndPunctfalse
  {\let\EndOfBibitem\relax}
\providecommand*\mciteSetBstMidEndSepPunct[3]{}
\providecommand*\mciteSetBstSublistLabelBeginEnd[3]{}
\providecommand*\EndOfBibitem{}
\mciteSetBstSublistMode{f}
\mciteSetBstMaxWidthForm{subitem}{(\alph{mcitesubitemcount})}
\mciteSetBstSublistLabelBeginEnd
  {\mcitemaxwidthsubitemform\space}
  {\relax}
  {\relax}

\bibitem[Kreuer(1996)]{kreuer1996proton}
Kreuer,~K.-D. Proton conductivity: Materials and applications. \emph{Chemistry
  of materials} \textbf{1996}, \emph{8}, 610--641\relax
\mciteBstWouldAddEndPuncttrue
\mciteSetBstMidEndSepPunct{\mcitedefaultmidpunct}
{\mcitedefaultendpunct}{\mcitedefaultseppunct}\relax
\EndOfBibitem
\bibitem[Cukierman(2006)]{cukierman2006et}
Cukierman,~S. Et tu, Grotthuss! and other unfinished stories. \emph{Biochimica
  et Biophysica Acta (BBA)-Biomembranes} \textbf{2006}, \emph{1757},
  876--885\relax
\mciteBstWouldAddEndPuncttrue
\mciteSetBstMidEndSepPunct{\mcitedefaultmidpunct}
{\mcitedefaultendpunct}{\mcitedefaultseppunct}\relax
\EndOfBibitem
\bibitem[Wraight(2006)]{wraight2006chance}
Wraight,~C.~A. Chance and design - Proton transfer in water, channels and
  bioenergetic proteins. \emph{Biochimica et Biophysica Acta
  (BBA)-Biomembranes} \textbf{2006}, \emph{1757}, 886--912\relax
\mciteBstWouldAddEndPuncttrue
\mciteSetBstMidEndSepPunct{\mcitedefaultmidpunct}
{\mcitedefaultendpunct}{\mcitedefaultseppunct}\relax
\EndOfBibitem
\bibitem[Raymand \latin{et~al.}(2011)Raymand, van Duin, Goddard, Hermansson,
  and Spångberg]{raymond2011}
Raymand,~D.; van Duin,~A.~C.; Goddard,~W. A.~I.; Hermansson,~K.; Spångberg,~D.
  Hydroxylation Structure and Proton Transfer Reactivity at the Zinc
  Oxide-Water Interface. \emph{The Journal of Physical Chemistry C}
  \textbf{2011}, \emph{115}, 8573--8579\relax
\mciteBstWouldAddEndPuncttrue
\mciteSetBstMidEndSepPunct{\mcitedefaultmidpunct}
{\mcitedefaultendpunct}{\mcitedefaultseppunct}\relax
\EndOfBibitem
\bibitem[Yamaguchi \latin{et~al.}(2014)Yamaguchi, Inuzuka, Takashima, Hayashi,
  Hashimoto, and Nakamura]{Yamaguchi2014}
Yamaguchi,~A.; Inuzuka,~R.; Takashima,~T.; Hayashi,~T.; Hashimoto,~K.;
  Nakamura,~R. Regulating proton-coupled electron transfer for efficient water
  splitting by manganese oxides at neutral pH. \emph{Nature Communications}
  \textbf{2014}, \emph{5}, 4256\relax
\mciteBstWouldAddEndPuncttrue
\mciteSetBstMidEndSepPunct{\mcitedefaultmidpunct}
{\mcitedefaultendpunct}{\mcitedefaultseppunct}\relax
\EndOfBibitem
\bibitem[Ma \latin{et~al.}(2022)Ma, Shi, Liu, Li, Cui, Tan, Zhao, and
  Wang]{HBN_watersplitting}
Ma,~X.; Shi,~Y.; Liu,~J.; Li,~X.; Cui,~X.; Tan,~S.; Zhao,~J.; Wang,~B.
  Hydrogen-Bond Network Promotes Water Splitting on the TiO2 Surface.
  \emph{Journal of the American Chemical Society} \textbf{2022}, \emph{144},
  13565--13573, PMID: 35852138\relax
\mciteBstWouldAddEndPuncttrue
\mciteSetBstMidEndSepPunct{\mcitedefaultmidpunct}
{\mcitedefaultendpunct}{\mcitedefaultseppunct}\relax
\EndOfBibitem
\bibitem[Fajín \latin{et~al.}(2014)Fajín, D.~S.~Cordeiro, and
  Gomes]{gomes2014}
Fajín,~J. L.~C.; D.~S.~Cordeiro,~M.~N.; Gomes,~J. R.~B. Density Functional
  Theory Study of the Water Dissociation on Platinum Surfaces: General Trends.
  \emph{The Journal of Physical Chemistry A} \textbf{2014}, \emph{118},
  5832--5840, PMID: 24547954\relax
\mciteBstWouldAddEndPuncttrue
\mciteSetBstMidEndSepPunct{\mcitedefaultmidpunct}
{\mcitedefaultendpunct}{\mcitedefaultseppunct}\relax
\EndOfBibitem
\bibitem[Warburton \latin{et~al.}(2021)Warburton, Mayer, and
  Hammes-Schiffer]{Warburton2021}
Warburton,~R.~E.; Mayer,~J.~M.; Hammes-Schiffer,~S. Proton-Coupled Defects
  Impact O–H Bond Dissociation Free Energies on Metal Oxide Surfaces.
  \emph{The Journal of Physical Chemistry Letters} \textbf{2021}, \emph{12},
  9761--9767, PMID: 34595925\relax
\mciteBstWouldAddEndPuncttrue
\mciteSetBstMidEndSepPunct{\mcitedefaultmidpunct}
{\mcitedefaultendpunct}{\mcitedefaultseppunct}\relax
\EndOfBibitem
\bibitem[Creazzo and Luber(2023)Creazzo, and Luber]{Creazzo2023}
Creazzo,~F.; Luber,~S. \emph{Reference Module in Chemistry, Molecular Sciences
  and Chemical Engineering}; Elsevier, 2023\relax
\mciteBstWouldAddEndPuncttrue
\mciteSetBstMidEndSepPunct{\mcitedefaultmidpunct}
{\mcitedefaultendpunct}{\mcitedefaultseppunct}\relax
\EndOfBibitem
\bibitem[Agmon \latin{et~al.}(2016)Agmon, Bakker, Campen, Henchman, Pohl, Roke,
  Th\"amer, and Hassanali]{ChemRev3}
Agmon,~N.; Bakker,~H.~J.; Campen,~R.~K.; Henchman,~R.~H.; Pohl,~P.; Roke,~S.;
  Th\"amer,~M.; Hassanali,~A. Protons and Hydroxide Ions in Aqueous Systems.
  \emph{Chemical Reviews} \textbf{2016}, \emph{116}, 7642--7672\relax
\mciteBstWouldAddEndPuncttrue
\mciteSetBstMidEndSepPunct{\mcitedefaultmidpunct}
{\mcitedefaultendpunct}{\mcitedefaultseppunct}\relax
\EndOfBibitem
\bibitem[Stuchebrukhov(2009)]{stuchebrukov}
Stuchebrukhov,~A.~A. Mechanisms of proton transfer in proteins: Localized
  charge transfer versus delocalized soliton transfer. \emph{Phys. Rev. E}
  \textbf{2009}, \emph{79}, 031927\relax
\mciteBstWouldAddEndPuncttrue
\mciteSetBstMidEndSepPunct{\mcitedefaultmidpunct}
{\mcitedefaultendpunct}{\mcitedefaultseppunct}\relax
\EndOfBibitem
\bibitem[Persson and Halle(2015)Persson, and Halle]{halle_pnas2015}
Persson,~F.; Halle,~B. How amide hydrogens exchange in native proteins.
  \emph{Proceedings of the National Academy of Sciences} \textbf{2015},
  \emph{112}, 10383--10388\relax
\mciteBstWouldAddEndPuncttrue
\mciteSetBstMidEndSepPunct{\mcitedefaultmidpunct}
{\mcitedefaultendpunct}{\mcitedefaultseppunct}\relax
\EndOfBibitem
\bibitem[Li and Voth(2021)Li, and Voth]{livoth2021}
Li,~C.; Voth,~G.~A. A quantitative paradigm for water-assisted proton transport
  through proteins and other confined spaces. \emph{Proceedings of the National
  Academy of Sciences} \textbf{2021}, \emph{118}, e2113141118\relax
\mciteBstWouldAddEndPuncttrue
\mciteSetBstMidEndSepPunct{\mcitedefaultmidpunct}
{\mcitedefaultendpunct}{\mcitedefaultseppunct}\relax
\EndOfBibitem
\bibitem[Hynes(1999)]{Hynes1997}
Hynes,~J.~T. Physical chemistry: The protean proton in water. \emph{Nature}
  \textbf{1999}, \emph{397}, 565--567\relax
\mciteBstWouldAddEndPuncttrue
\mciteSetBstMidEndSepPunct{\mcitedefaultmidpunct}
{\mcitedefaultendpunct}{\mcitedefaultseppunct}\relax
\EndOfBibitem
\bibitem[de~Grotthuss(1806)]{Grotthuss1806}
de~Grotthuss,~C. J.~T. Theory of decomposition of liquids by electrical
  currents. \emph{Ann. Chim. (Paris)} \textbf{1806}, \emph{LVIII}, 54--74\relax
\mciteBstWouldAddEndPuncttrue
\mciteSetBstMidEndSepPunct{\mcitedefaultmidpunct}
{\mcitedefaultendpunct}{\mcitedefaultseppunct}\relax
\EndOfBibitem
\bibitem[Agmon(1995)]{Agmon1995}
Agmon,~N. The {G}rotthuss mechanism. \emph{Chemical Physics Letters}
  \textbf{1995}, \emph{244}, 456 -- 462\relax
\mciteBstWouldAddEndPuncttrue
\mciteSetBstMidEndSepPunct{\mcitedefaultmidpunct}
{\mcitedefaultendpunct}{\mcitedefaultseppunct}\relax
\EndOfBibitem
\bibitem[Atkins and de~Paula(2006)Atkins, and de~Paula]{Atkins}
Atkins,~P.; de~Paula,~J. \emph{Physical Chemistry, Eight Edition}; Oxford
  University Press, 2006\relax
\mciteBstWouldAddEndPuncttrue
\mciteSetBstMidEndSepPunct{\mcitedefaultmidpunct}
{\mcitedefaultendpunct}{\mcitedefaultseppunct}\relax
\EndOfBibitem
\bibitem[Eigen and De~Maeyer(1958)Eigen, and De~Maeyer]{eigen1958self}
Eigen,~M.; De~Maeyer,~L. Self-dissociation and protonic charge transport in
  water and ice. \emph{Proceedings of the Royal Society of London. Series A.
  Mathematical and Physical Sciences} \textbf{1958}, \emph{247}, 505--533\relax
\mciteBstWouldAddEndPuncttrue
\mciteSetBstMidEndSepPunct{\mcitedefaultmidpunct}
{\mcitedefaultendpunct}{\mcitedefaultseppunct}\relax
\EndOfBibitem
\bibitem[Zundel(1969)]{zundel1969hydration}
Zundel,~G. Hydration structure and intermolecular interaction in
  polyelectrolytes. \emph{Angewandte Chemie International Edition in English}
  \textbf{1969}, \emph{8}, 499--509\relax
\mciteBstWouldAddEndPuncttrue
\mciteSetBstMidEndSepPunct{\mcitedefaultmidpunct}
{\mcitedefaultendpunct}{\mcitedefaultseppunct}\relax
\EndOfBibitem
\bibitem[Eigen(1964)]{eigen1964proton}
Eigen,~M. Proton transfer, acid-base catalysis, and enzymatic hydrolysis. I.
  Elementary processes. \emph{Angewandte Chemie International Edition in
  English} \textbf{1964}, \emph{3}, 1--499\relax
\mciteBstWouldAddEndPuncttrue
\mciteSetBstMidEndSepPunct{\mcitedefaultmidpunct}
{\mcitedefaultendpunct}{\mcitedefaultseppunct}\relax
\EndOfBibitem
\bibitem[Danninger and Zundel(1981)Danninger, and Zundel]{zundel1981}
Danninger,~W.; Zundel,~G. Intense Depolarized Rayleigh-Scattering in
  Raman-Spectra of Acids Caused by Large Proton Polarizabilities of
  Hydrogen-Bonds. \emph{J. Chem. Phys.} \textbf{1981}, \emph{74}, 2769\relax
\mciteBstWouldAddEndPuncttrue
\mciteSetBstMidEndSepPunct{\mcitedefaultmidpunct}
{\mcitedefaultendpunct}{\mcitedefaultseppunct}\relax
\EndOfBibitem
\bibitem[Zundel(2000)]{Zundel2000}
Zundel,~G. Hydrogen Bonds with Large Proton Polarizability and Proton Transfer
  Processes in Electrochemistry and Biology. \emph{Adv. Chem. Phys.}
  \textbf{2000}, \emph{111}, 1\relax
\mciteBstWouldAddEndPuncttrue
\mciteSetBstMidEndSepPunct{\mcitedefaultmidpunct}
{\mcitedefaultendpunct}{\mcitedefaultseppunct}\relax
\EndOfBibitem
\bibitem[Botti \latin{et~al.}(2006)Botti, Bruni, Ricci, and Soper]{botti2006}
Botti,~A.; Bruni,~F.; Ricci,~M.~A.; Soper,~A.~K. Eigen Versus Zundel Complexes
  in Hcl-Water Mixtures. \emph{J. Chem. Phys.} \textbf{2006}, \emph{125},
  014508\relax
\mciteBstWouldAddEndPuncttrue
\mciteSetBstMidEndSepPunct{\mcitedefaultmidpunct}
{\mcitedefaultendpunct}{\mcitedefaultseppunct}\relax
\EndOfBibitem
\bibitem[Tielrooij \latin{et~al.}(2009)Tielrooij, Timmer, Bakker, and
  Bonn]{bakkerbonn2009}
Tielrooij,~K.~J.; Timmer,~R. L.~A.; Bakker,~H.~J.; Bonn,~M. Structure Dynamics
  of the Proton in Liquid Water Probed with Terahertz Time-Domain Spectroscopy.
  \emph{Phys. Rev. Lett.} \textbf{2009}, \emph{102}, 198303\relax
\mciteBstWouldAddEndPuncttrue
\mciteSetBstMidEndSepPunct{\mcitedefaultmidpunct}
{\mcitedefaultendpunct}{\mcitedefaultseppunct}\relax
\EndOfBibitem
\bibitem[Thamer \latin{et~al.}(2015)Thamer, De~Marco, Ramasesha, Mandal, and
  Tokmakoff]{tokmakoff2015}
Thamer,~M.; De~Marco,~L.; Ramasesha,~K.; Mandal,~A.; Tokmakoff,~A. Ultrafast 2d
  Ir Spectroscopy of the Excess Proton in Liquid Water. \emph{Science}
  \textbf{2015}, \emph{350}, 78\relax
\mciteBstWouldAddEndPuncttrue
\mciteSetBstMidEndSepPunct{\mcitedefaultmidpunct}
{\mcitedefaultendpunct}{\mcitedefaultseppunct}\relax
\EndOfBibitem
\bibitem[Dahms \latin{et~al.}(2016)Dahms, Costard, Pines, Fingerhut, Nibbering,
  and Elsaesser]{dahms2016}
Dahms,~F.; Costard,~R.; Pines,~E.; Fingerhut,~B.~P.; Nibbering,~E. T.~J.;
  Elsaesser,~T. The Hydrated Excess Proton in the Zundel Cation H5O2+: The Role
  of Ultrafast Solvent Fluctuations. \emph{Angewandte Chemie International
  Edition} \textbf{2016}, \emph{55}, 10600--10605\relax
\mciteBstWouldAddEndPuncttrue
\mciteSetBstMidEndSepPunct{\mcitedefaultmidpunct}
{\mcitedefaultendpunct}{\mcitedefaultseppunct}\relax
\EndOfBibitem
\bibitem[Kozari \latin{et~al.}(2021)Kozari, Sigalov, Pines, Fingerhut, and
  Pines]{kozari2021}
Kozari,~E.; Sigalov,~M.; Pines,~D.; Fingerhut,~B.~P.; Pines,~E. Infrared and
  NMR Spectroscopic Fingerprints of the Asymmetric H7+O3 Complex in Solution.
  \emph{ChemPhysChem} \textbf{2021}, \emph{22}, 716--725\relax
\mciteBstWouldAddEndPuncttrue
\mciteSetBstMidEndSepPunct{\mcitedefaultmidpunct}
{\mcitedefaultendpunct}{\mcitedefaultseppunct}\relax
\EndOfBibitem
\bibitem[Hückel(1928)]{huckel1928}
Hückel,~Z. Theorie der Beweglichkeiten des Wasserstoff- und Hydroxylions in
  wässeriger Lösung. \emph{Zeitschrift für Elektrochemie} \textbf{1928},
  \emph{34}, 546--562\relax
\mciteBstWouldAddEndPuncttrue
\mciteSetBstMidEndSepPunct{\mcitedefaultmidpunct}
{\mcitedefaultendpunct}{\mcitedefaultseppunct}\relax
\EndOfBibitem
\bibitem[Tuckerman \latin{et~al.}(1997)Tuckerman, Marx, Klein, and
  Parrinello]{TuckermanMarxKleinParrinello1997}
Tuckerman,~M.~E.; Marx,~D.; Klein,~M.~L.; Parrinello,~M. On the quantum nature
  of the shared proton in hydrogen bonds. \emph{Science} \textbf{1997},
  \emph{275}, 817\relax
\mciteBstWouldAddEndPuncttrue
\mciteSetBstMidEndSepPunct{\mcitedefaultmidpunct}
{\mcitedefaultendpunct}{\mcitedefaultseppunct}\relax
\EndOfBibitem
\bibitem[Marx \latin{et~al.}(1999)Marx, Tuckerman, Hutter, and
  Parrinello]{MarxTuckermanHutterParrinello1999}
Marx,~D.; Tuckerman,~M.~E.; Hutter,~J.; Parrinello,~M. The nature of the
  hydrated excess proton in water. \emph{Nature} \textbf{1999}, \emph{397},
  601--604\relax
\mciteBstWouldAddEndPuncttrue
\mciteSetBstMidEndSepPunct{\mcitedefaultmidpunct}
{\mcitedefaultendpunct}{\mcitedefaultseppunct}\relax
\EndOfBibitem
\bibitem[Marx \latin{et~al.}(2010)Marx, Chandra, and
  Tuckerman]{marx2010aqueous}
Marx,~D.; Chandra,~A.; Tuckerman,~M.~E. Aqueous basic solutions: Hydroxide
  solvation, structural diffusion, and comparison to the hydrated proton.
  \emph{Chemical reviews} \textbf{2010}, \emph{110}, 2174--2216\relax
\mciteBstWouldAddEndPuncttrue
\mciteSetBstMidEndSepPunct{\mcitedefaultmidpunct}
{\mcitedefaultendpunct}{\mcitedefaultseppunct}\relax
\EndOfBibitem
\bibitem[Tse \latin{et~al.}(2015)Tse, Knight, and Voth]{tse_analysis_2015}
Tse,~Y.-L.~S.; Knight,~C.; Voth,~G.~A. An analysis of hydrated proton diffusion
  in ab initio molecular dynamics. \emph{The Journal of Chemical Physics}
  \textbf{2015}, \emph{142}, 014104, \_eprint:
  https://pubs.aip.org/aip/jcp/article-pdf/doi/10.1063/1.4905077/13607981/014104\_1\_online.pdf\relax
\mciteBstWouldAddEndPuncttrue
\mciteSetBstMidEndSepPunct{\mcitedefaultmidpunct}
{\mcitedefaultendpunct}{\mcitedefaultseppunct}\relax
\EndOfBibitem
\bibitem[Hassanali \latin{et~al.}(2013)Hassanali, Giberti, Cuny, Kühne, and
  Parrinello]{HassanaliGibertiCunyKuhneParrinello2013}
Hassanali,~A.; Giberti,~F.; Cuny,~J.; Kühne,~T.~D.; Parrinello,~M. Proton
  transfer through the water gossamer. \emph{Proceedings of the National
  Academy of Sciences} \textbf{2013}, 13723\relax
\mciteBstWouldAddEndPuncttrue
\mciteSetBstMidEndSepPunct{\mcitedefaultmidpunct}
{\mcitedefaultendpunct}{\mcitedefaultseppunct}\relax
\EndOfBibitem
\bibitem[Hassanali \latin{et~al.}(2014)Hassanali, Giberti, Sosso, and
  Parrinello]{HassanaliGibertiSossoParrinello2014}
Hassanali,~A.~A.; Giberti,~F.; Sosso,~G.~C.; Parrinello,~M. The role of the
  umbrella inversion mode in proton diffusion. \emph{Chemical Physics Letters}
  \textbf{2014}, \emph{599}, 133 -- 138\relax
\mciteBstWouldAddEndPuncttrue
\mciteSetBstMidEndSepPunct{\mcitedefaultmidpunct}
{\mcitedefaultendpunct}{\mcitedefaultseppunct}\relax
\EndOfBibitem
\bibitem[Cassone(2020)]{cassone2020}
Cassone,~G. Nuclear Quantum Effects Largely Influence Molecular Dissociation
  and Proton Transfer in Liquid Water under an Electric Field. \emph{The
  Journal of Physical Chemistry Letters} \textbf{2020}, \emph{11}, 8983--8988,
  PMID: 33035059\relax
\mciteBstWouldAddEndPuncttrue
\mciteSetBstMidEndSepPunct{\mcitedefaultmidpunct}
{\mcitedefaultendpunct}{\mcitedefaultseppunct}\relax
\EndOfBibitem
\bibitem[Lobaugh and Voth(1996)Lobaugh, and Voth]{LobaughVoth1996}
Lobaugh,~J.; Voth,~G.~A. The quantum dynamics of an excess proton in water.
  \emph{The Journal of Chemical Physics} \textbf{1996}, \emph{104},
  2056--2069\relax
\mciteBstWouldAddEndPuncttrue
\mciteSetBstMidEndSepPunct{\mcitedefaultmidpunct}
{\mcitedefaultendpunct}{\mcitedefaultseppunct}\relax
\EndOfBibitem
\bibitem[Pavese and Voth(1998)Pavese, and Voth]{PaveseVoth1998}
Pavese,~M.; Voth,~G.~A. Quantum and classical simulations of an excess proton
  in water. \emph{Berichte der Bunsengesellschaft für physikalische Chemie}
  \textbf{1998}, \emph{102}, 527--532\relax
\mciteBstWouldAddEndPuncttrue
\mciteSetBstMidEndSepPunct{\mcitedefaultmidpunct}
{\mcitedefaultendpunct}{\mcitedefaultseppunct}\relax
\EndOfBibitem
\bibitem[Lapid \latin{et~al.}(2005)Lapid, Agmon, Petersen, and
  Voth]{LapidAgmonPetersenVoth2005}
Lapid,~H.; Agmon,~N.; Petersen,~M.~K.; Voth,~G.~A. A bond-order analysis of the
  mechanism for hydrated proton mobility in liquid water. \emph{The Journal of
  Chemical Physics} \textbf{2005}, \emph{122}, 014506\relax
\mciteBstWouldAddEndPuncttrue
\mciteSetBstMidEndSepPunct{\mcitedefaultmidpunct}
{\mcitedefaultendpunct}{\mcitedefaultseppunct}\relax
\EndOfBibitem
\bibitem[Wu \latin{et~al.}(2008)Wu, Chen, Wang, Paesani, and
  Voth]{WuChenWangPaesaniVoth2008}
Wu,~Y.; Chen,~H.; Wang,~F.; Paesani,~F.; Voth,~G.~A. An Improved Multistate
  Empirical Valence Bond Model for Aqueous Proton Solvation and Transport.
  \emph{The Journal of Physical Chemistry B} \textbf{2008}, \emph{112},
  467--482, PMID: 17999484\relax
\mciteBstWouldAddEndPuncttrue
\mciteSetBstMidEndSepPunct{\mcitedefaultmidpunct}
{\mcitedefaultendpunct}{\mcitedefaultseppunct}\relax
\EndOfBibitem
\bibitem[Knight and Voth(2012)Knight, and Voth]{KnightVoth2012}
Knight,~C.; Voth,~G.~A. The Curious Case of the Hydrated Proton. \emph{Accounts
  of Chemical Research} \textbf{2012}, \emph{45}, 101--109\relax
\mciteBstWouldAddEndPuncttrue
\mciteSetBstMidEndSepPunct{\mcitedefaultmidpunct}
{\mcitedefaultendpunct}{\mcitedefaultseppunct}\relax
\EndOfBibitem
\bibitem[Tse \latin{et~al.}(2015)Tse, Chen, Lindberg, Kumar, and
  Voth]{TseLindbergKumarVoth2015}
Tse,~Y.-L.~S.; Chen,~C.; Lindberg,~G.~E.; Kumar,~R.; Voth,~G.~A. Propensity of
  Hydrated Excess Protons and Hydroxide Anions for the Air--Water Interface.
  \emph{Journal of the American Chemical Society} \textbf{2015}, \emph{137},
  12610--12616\relax
\mciteBstWouldAddEndPuncttrue
\mciteSetBstMidEndSepPunct{\mcitedefaultmidpunct}
{\mcitedefaultendpunct}{\mcitedefaultseppunct}\relax
\EndOfBibitem
\bibitem[Headrick \latin{et~al.}(2005)Headrick, Diken, Walters, Hammer,
  Christie, Cui, Myshakin, Duncan, Johnson, and Jordan]{headrick2005spectral}
Headrick,~J.~M.; Diken,~E.~G.; Walters,~R.~S.; Hammer,~N.~I.; Christie,~R.~A.;
  Cui,~J.; Myshakin,~E.~M.; Duncan,~M.~A.; Johnson,~M.~A.; Jordan,~K.~D.
  Spectral signatures of hydrated proton vibrations in water clusters.
  \emph{Science} \textbf{2005}, \emph{308}, 1765--1769\relax
\mciteBstWouldAddEndPuncttrue
\mciteSetBstMidEndSepPunct{\mcitedefaultmidpunct}
{\mcitedefaultendpunct}{\mcitedefaultseppunct}\relax
\EndOfBibitem
\bibitem[Yu and Cui(2007)Yu, and Cui]{yu2007vibrational}
Yu,~H.; Cui,~Q. The vibrational spectra of protonated water clusters: A
  benchmark for self-consistent-charge density-functional tight binding.
  \emph{The Journal of chemical physics} \textbf{2007}, \emph{127},
  234504\relax
\mciteBstWouldAddEndPuncttrue
\mciteSetBstMidEndSepPunct{\mcitedefaultmidpunct}
{\mcitedefaultendpunct}{\mcitedefaultseppunct}\relax
\EndOfBibitem
\bibitem[Park \latin{et~al.}(2007)Park, Shin, Singh, and Kim]{park2007eigen}
Park,~M.; Shin,~I.-H.; Singh,~N.~J.; Kim,~K.~S. Eigen and Zundel forms of small
  protonated water clusters: Structures and infrared spectra. \emph{The Journal
  of Physical Chemistry A} \textbf{2007}, \emph{111}, 10692--10702\relax
\mciteBstWouldAddEndPuncttrue
\mciteSetBstMidEndSepPunct{\mcitedefaultmidpunct}
{\mcitedefaultendpunct}{\mcitedefaultseppunct}\relax
\EndOfBibitem
\bibitem[Dong and Nesbitt(2006)Dong, and Nesbitt]{DongNesbitt2006}
Dong,~F.; Nesbitt,~D.~J. Jet cooled spectroscopy of $H_2DO^+$: Barrier heights
  and isotope-dependent tunneling dynamics from $H_3O^+$ to $D_3O^+$. \emph{The
  Journal of Chemical Physics} \textbf{2006}, \emph{125}, 144311\relax
\mciteBstWouldAddEndPuncttrue
\mciteSetBstMidEndSepPunct{\mcitedefaultmidpunct}
{\mcitedefaultendpunct}{\mcitedefaultseppunct}\relax
\EndOfBibitem
\bibitem[Sobolewski and Domcke(2002)Sobolewski, and
  Domcke]{SobolweskiDomcke2002}
Sobolewski,~A.~L.; Domcke,~W. Ab Initio Investigation of the Structure and
  Spectroscopy of Hydronium-Water Clusters. \emph{The Journal of Physical
  Chemistry A} \textbf{2002}, \emph{106}, 4158--4167\relax
\mciteBstWouldAddEndPuncttrue
\mciteSetBstMidEndSepPunct{\mcitedefaultmidpunct}
{\mcitedefaultendpunct}{\mcitedefaultseppunct}\relax
\EndOfBibitem
\bibitem[Fournier \latin{et~al.}(2015)Fournier, Wolke, Johnson, Odbadrakh,
  Jordan, Kathmann, and Xantheas]{fournier2015snapshots}
Fournier,~J.~A.; Wolke,~C.~T.; Johnson,~M.~A.; Odbadrakh,~T.~T.; Jordan,~K.~D.;
  Kathmann,~S.~M.; Xantheas,~S.~S. Snapshots of proton accommodation at a
  microscopic water surface: Understanding the vibrational spectral signatures
  of the charge defect in cryogenically cooled H+(H2O)n=2-28 clusters.
  \emph{The Journal of Physical Chemistry A} \textbf{2015}, \emph{119},
  9425--9440\relax
\mciteBstWouldAddEndPuncttrue
\mciteSetBstMidEndSepPunct{\mcitedefaultmidpunct}
{\mcitedefaultendpunct}{\mcitedefaultseppunct}\relax
\EndOfBibitem
\bibitem[Fournier \latin{et~al.}(2014)Fournier, Wolke, Johnson, Johnson, Heine,
  Gewinner, Schöllkopf, Esser, Fagiani, Knorke, and Asmis]{fournier2014}
Fournier,~J.~A.; Wolke,~C.~T.; Johnson,~C.~J.; Johnson,~M.~A.; Heine,~N.;
  Gewinner,~S.; Schöllkopf,~W.; Esser,~T.~K.; Fagiani,~M.~R.; Knorke,~H.;
  Asmis,~K.~R. Site-specific vibrational spectral signatures of water molecules
  in the magic $H_3O^+(H_2O)_{20}$ and $Cs^+(H_2O)_{20}$ clusters.
  \emph{Proceedings of the National Academy of Sciences} \textbf{2014},
  \emph{111}, 18132--18137\relax
\mciteBstWouldAddEndPuncttrue
\mciteSetBstMidEndSepPunct{\mcitedefaultmidpunct}
{\mcitedefaultendpunct}{\mcitedefaultseppunct}\relax
\EndOfBibitem
\bibitem[Shin \latin{et~al.}(2004)Shin, Hammer, Diken, Johnson, Walters,
  Jaeger, Duncan, Christie, and Jordan]{jordan2004}
Shin,~J.-W.; Hammer,~N.~I.; Diken,~E.~G.; Johnson,~M.~A.; Walters,~R.~S.;
  Jaeger,~T.~D.; Duncan,~M.~A.; Christie,~R.~A.; Jordan,~K.~D. Infrared
  Signature of Structures Associated with the $H^+(H_2O)_n$ (n = 6 to 27)
  Clusters. \emph{Science} \textbf{2004}, \emph{304}, 1137--1140\relax
\mciteBstWouldAddEndPuncttrue
\mciteSetBstMidEndSepPunct{\mcitedefaultmidpunct}
{\mcitedefaultendpunct}{\mcitedefaultseppunct}\relax
\EndOfBibitem
\bibitem[Calio \latin{et~al.}(2021)Calio, Li, and Voth]{vothcalio2021}
Calio,~P.~B.; Li,~C.; Voth,~G.~A. Resolving the Structural Debate for the
  Hydrated Excess Proton in Water. \emph{Journal of the American Chemical
  Society} \textbf{2021}, \emph{143}, 18672--18683, PMID: 34723507\relax
\mciteBstWouldAddEndPuncttrue
\mciteSetBstMidEndSepPunct{\mcitedefaultmidpunct}
{\mcitedefaultendpunct}{\mcitedefaultseppunct}\relax
\EndOfBibitem
\bibitem[Fournier \latin{et~al.}(2018)Fournier, Carpenter, Lewis, and
  Tokmakoff]{fournierbroadband_2018}
Fournier,~J.~A.; Carpenter,~W.~B.; Lewis,~N. H.~C.; Tokmakoff,~A. Broadband
  {2D} {IR} spectroscopy reveals dominant asymmetric {H5O2}+ proton hydration
  structures in acid solutions. \emph{Nature Chemistry} \textbf{2018},
  \emph{10}, 932--937\relax
\mciteBstWouldAddEndPuncttrue
\mciteSetBstMidEndSepPunct{\mcitedefaultmidpunct}
{\mcitedefaultendpunct}{\mcitedefaultseppunct}\relax
\EndOfBibitem
\bibitem[Carpenter \latin{et~al.}(2018)Carpenter, Fournier, Lewis, and
  Tokmakoff]{william2018}
Carpenter,~W.~B.; Fournier,~J.~A.; Lewis,~N. H.~C.; Tokmakoff,~A. Picosecond
  Proton Transfer Kinetics in Water Revealed with Ultrafast IR Spectroscopy.
  \emph{The Journal of Physical Chemistry B} \textbf{2018}, \emph{122},
  2792--2802, PMID: 29452488\relax
\mciteBstWouldAddEndPuncttrue
\mciteSetBstMidEndSepPunct{\mcitedefaultmidpunct}
{\mcitedefaultendpunct}{\mcitedefaultseppunct}\relax
\EndOfBibitem
\bibitem[Carpenter \latin{et~al.}(2019)Carpenter, Lewis, Fournier, and
  Tokmakoff]{carpenter_entropic_2019}
Carpenter,~W.~B.; Lewis,~N. H.~C.; Fournier,~J.~A.; Tokmakoff,~A. Entropic
  barriers in the kinetics of aqueous proton transfer. \emph{The Journal of
  Chemical Physics} \textbf{2019}, \emph{151}, 034501, \_eprint:
  https://pubs.aip.org/aip/jcp/article-pdf/doi/10.1063/1.5108907/13941564/034501\_1\_online.pdf\relax
\mciteBstWouldAddEndPuncttrue
\mciteSetBstMidEndSepPunct{\mcitedefaultmidpunct}
{\mcitedefaultendpunct}{\mcitedefaultseppunct}\relax
\EndOfBibitem
\bibitem[Vuilleumier and Borgis(1998)Vuilleumier, and
  Borgis]{vuilleumier1998quantum}
Vuilleumier,~R.; Borgis,~D. Quantum Dynamics of an Excess Proton in Water Using
  an Extended Empirical Valence-Bond Hamiltonian. \emph{Journal of Physical
  Chemistry B} \textbf{1998}, \emph{102}, 4261--4264\relax
\mciteBstWouldAddEndPuncttrue
\mciteSetBstMidEndSepPunct{\mcitedefaultmidpunct}
{\mcitedefaultendpunct}{\mcitedefaultseppunct}\relax
\EndOfBibitem
\bibitem[Vuilleumier and Borgis(1999)Vuilleumier, and
  Borgis]{vuilleumier1999extended}
Vuilleumier,~R.; Borgis,~D. An Extended Empirical Valence Bond Model for
  Describing Proton Mobility in Water. \emph{Israel Journal of Chemistry}
  \textbf{1999}, \emph{39}, 457--467\relax
\mciteBstWouldAddEndPuncttrue
\mciteSetBstMidEndSepPunct{\mcitedefaultmidpunct}
{\mcitedefaultendpunct}{\mcitedefaultseppunct}\relax
\EndOfBibitem
\bibitem[Glielmo \latin{et~al.}(2021)Glielmo, Husic, Rodriguez, Clementi,
  No{\'e}, and Laio]{glielmo2021unsupervised}
Glielmo,~A.; Husic,~B.~E.; Rodriguez,~A.; Clementi,~C.; No{\'e},~F.; Laio,~A.
  Unsupervised Learning Methods for Molecular Simulation Data. \emph{Chemical
  Reviews} \textbf{2021}, \emph{121}, 9722--9758\relax
\mciteBstWouldAddEndPuncttrue
\mciteSetBstMidEndSepPunct{\mcitedefaultmidpunct}
{\mcitedefaultendpunct}{\mcitedefaultseppunct}\relax
\EndOfBibitem
\bibitem[Glielmo \latin{et~al.}(2022)Glielmo, Zeni, Cheng, Cs{\'a}nyi, and
  Laio]{glielmo2022ranking}
Glielmo,~A.; Zeni,~C.; Cheng,~B.; Cs{\'a}nyi,~G.; Laio,~A. Ranking the
  information content of distance measures. \emph{PNAS Nexus} \textbf{2022},
  \emph{1}, pgac039\relax
\mciteBstWouldAddEndPuncttrue
\mciteSetBstMidEndSepPunct{\mcitedefaultmidpunct}
{\mcitedefaultendpunct}{\mcitedefaultseppunct}\relax
\EndOfBibitem
\bibitem[Offei-Danso \latin{et~al.}(2022)Offei-Danso, Hassanali, and
  Rodriguez]{offei2022high}
Offei-Danso,~A.; Hassanali,~A.; Rodriguez,~A. High-Dimensional Fluctuations in
  Liquid Water: Combining Chemical Intuition with Unsupervised Learning.
  \emph{Journal of Chemical Theory and Computation} \textbf{2022}, \relax
\mciteBstWouldAddEndPunctfalse
\mciteSetBstMidEndSepPunct{\mcitedefaultmidpunct}
{}{\mcitedefaultseppunct}\relax
\EndOfBibitem
\bibitem[Napoli \latin{et~al.}(2018)Napoli, Marsalek, and Markland]{napoli2018}
Napoli,~J.~A.; Marsalek,~O.; Markland,~T.~E. {Decoding the spectroscopic
  features and time scales of aqueous proton defects}. \emph{The Journal of
  Chemical Physics} \textbf{2018}, \emph{148}, 222833\relax
\mciteBstWouldAddEndPuncttrue
\mciteSetBstMidEndSepPunct{\mcitedefaultmidpunct}
{\mcitedefaultendpunct}{\mcitedefaultseppunct}\relax
\EndOfBibitem
\bibitem[Bart{\'o}k \latin{et~al.}(2013)Bart{\'o}k, Kondor, and
  Cs{\'a}nyi]{bartok2013representing}
Bart{\'o}k,~A.~P.; Kondor,~R.; Cs{\'a}nyi,~G. On representing chemical
  environments. \emph{Physical Review B} \textbf{2013}, \emph{87}, 184115\relax
\mciteBstWouldAddEndPuncttrue
\mciteSetBstMidEndSepPunct{\mcitedefaultmidpunct}
{\mcitedefaultendpunct}{\mcitedefaultseppunct}\relax
\EndOfBibitem
\bibitem[Facco \latin{et~al.}(2017)Facco, d’Errico, Rodriguez, and
  Laio]{facco2017estimating}
Facco,~E.; d’Errico,~M.; Rodriguez,~A.; Laio,~A. Estimating the intrinsic
  dimension of datasets by a minimal neighborhood information. \emph{Scientific
  reports} \textbf{2017}, \emph{7}, 1--8\relax
\mciteBstWouldAddEndPuncttrue
\mciteSetBstMidEndSepPunct{\mcitedefaultmidpunct}
{\mcitedefaultendpunct}{\mcitedefaultseppunct}\relax
\EndOfBibitem
\bibitem[Rodriguez and Laio(2014)Rodriguez, and Laio]{rodriguez2014clustering}
Rodriguez,~A.; Laio,~A. Clustering by fast search and find of density peaks.
  \emph{science} \textbf{2014}, \emph{344}, 1492--1496\relax
\mciteBstWouldAddEndPuncttrue
\mciteSetBstMidEndSepPunct{\mcitedefaultmidpunct}
{\mcitedefaultendpunct}{\mcitedefaultseppunct}\relax
\EndOfBibitem
\bibitem[Rodriguez \latin{et~al.}(2018)Rodriguez, d’Errico, Facco, and
  Laio]{rodriguez2018computing}
Rodriguez,~A.; d’Errico,~M.; Facco,~E.; Laio,~A. Computing the free energy
  without collective variables. \emph{Journal of chemical theory and
  computation} \textbf{2018}, \emph{14}, 1206--1215\relax
\mciteBstWouldAddEndPuncttrue
\mciteSetBstMidEndSepPunct{\mcitedefaultmidpunct}
{\mcitedefaultendpunct}{\mcitedefaultseppunct}\relax
\EndOfBibitem
\bibitem[d'Errico \latin{et~al.}(2018)d'Errico, Facco, Laio, and
  Rodriguez]{d2018automatic}
d'Errico,~M.; Facco,~E.; Laio,~A.; Rodriguez,~A. Automatic topography of
  high-dimensional data sets by non-parametric Density Peak clustering.
  \emph{arXiv preprint arXiv:1802.10549} \textbf{2018}, \relax
\mciteBstWouldAddEndPunctfalse
\mciteSetBstMidEndSepPunct{\mcitedefaultmidpunct}
{}{\mcitedefaultseppunct}\relax
\EndOfBibitem
\bibitem[McInnes \latin{et~al.}(2018)McInnes, Healy, and
  Melville]{mcinnes2018umap}
McInnes,~L.; Healy,~J.; Melville,~J. Umap: Uniform manifold approximation and
  projection for dimension reduction. \emph{arXiv preprint arXiv:1802.03426}
  \textbf{2018}, \relax
\mciteBstWouldAddEndPunctfalse
\mciteSetBstMidEndSepPunct{\mcitedefaultmidpunct}
{}{\mcitedefaultseppunct}\relax
\EndOfBibitem
\bibitem[Perdew \latin{et~al.}(1996)Perdew, Burke, and Ernzerhof]{PBE}
Perdew,~J.~P.; Burke,~K.; Ernzerhof,~M. Generalized Gradient Approximation Made
  Simple. \emph{Phys. Rev. Lett.} \textbf{1996}, \emph{77}, 3865--3868\relax
\mciteBstWouldAddEndPuncttrue
\mciteSetBstMidEndSepPunct{\mcitedefaultmidpunct}
{\mcitedefaultendpunct}{\mcitedefaultseppunct}\relax
\EndOfBibitem
\bibitem[Zhang and Yang(1998)Zhang, and Yang]{revPBE}
Zhang,~Y.; Yang,~W. Comment on ``Generalized Gradient Approximation Made
  Simple''. \emph{Phys. Rev. Lett.} \textbf{1998}, \emph{80}, 890--890\relax
\mciteBstWouldAddEndPuncttrue
\mciteSetBstMidEndSepPunct{\mcitedefaultmidpunct}
{\mcitedefaultendpunct}{\mcitedefaultseppunct}\relax
\EndOfBibitem
\bibitem[Grimme \latin{et~al.}(2010)Grimme, Antony, Ehrlich, and
  Krieg]{grimme2010consistent}
Grimme,~S.; Antony,~J.; Ehrlich,~S.; Krieg,~H. A consistent and accurate ab
  initio parametrization of density functional dispersion correction (DFT-D)
  for the 94 elements H-Pu. \emph{The Journal of chemical physics}
  \textbf{2010}, \emph{132}, 154104\relax
\mciteBstWouldAddEndPuncttrue
\mciteSetBstMidEndSepPunct{\mcitedefaultmidpunct}
{\mcitedefaultendpunct}{\mcitedefaultseppunct}\relax
\EndOfBibitem
\bibitem[Maksimov \latin{et~al.}(2021)Maksimov, Baldauf, and
  Rossi]{maksimov2021conformational}
Maksimov,~D.; Baldauf,~C.; Rossi,~M. The conformational space of a flexible
  amino acid at metallic surfaces. \emph{International Journal of Quantum
  Chemistry} \textbf{2021}, \emph{121}, e26369\relax
\mciteBstWouldAddEndPuncttrue
\mciteSetBstMidEndSepPunct{\mcitedefaultmidpunct}
{\mcitedefaultendpunct}{\mcitedefaultseppunct}\relax
\EndOfBibitem
\bibitem[Grant(2020)]{grant2020network}
Grant,~W. Network modularity and local environment similarity as descriptors of
  protein structure. Ph.D.\ thesis, University of Cambridge, 2020\relax
\mciteBstWouldAddEndPuncttrue
\mciteSetBstMidEndSepPunct{\mcitedefaultmidpunct}
{\mcitedefaultendpunct}{\mcitedefaultseppunct}\relax
\EndOfBibitem
\bibitem[De \latin{et~al.}(2016)De, Bart{\'o}k, Cs{\'a}nyi, and
  Ceriotti]{de2016comparing}
De,~S.; Bart{\'o}k,~A.~P.; Cs{\'a}nyi,~G.; Ceriotti,~M. Comparing molecules and
  solids across structural and alchemical space. \emph{Physical Chemistry
  Chemical Physics} \textbf{2016}, \emph{18}, 13754--13769\relax
\mciteBstWouldAddEndPuncttrue
\mciteSetBstMidEndSepPunct{\mcitedefaultmidpunct}
{\mcitedefaultendpunct}{\mcitedefaultseppunct}\relax
\EndOfBibitem
\bibitem[Appignanesi \latin{et~al.}(2009)Appignanesi, Fris, and
  Sciortino]{appignanesi2009evidence}
Appignanesi,~G.~A.; Fris,~J.~R.; Sciortino,~F. Evidence of a two-state picture
  for supercooled water and its connections with glassy dynamics. \emph{The
  European Physical Journal E} \textbf{2009}, \emph{29}, 305--310\relax
\mciteBstWouldAddEndPuncttrue
\mciteSetBstMidEndSepPunct{\mcitedefaultmidpunct}
{\mcitedefaultendpunct}{\mcitedefaultseppunct}\relax
\EndOfBibitem
\bibitem[Capelli \latin{et~al.}(2022)Capelli, Muniz-Miranda, and
  Pavan]{capelli2022ephemeral}
Capelli,~R.; Muniz-Miranda,~F.; Pavan,~G.~M. Ephemeral ice-like local
  environments in classical rigid models of liquid water. \emph{The Journal of
  Chemical Physics} \textbf{2022}, \emph{156}, 214503\relax
\mciteBstWouldAddEndPuncttrue
\mciteSetBstMidEndSepPunct{\mcitedefaultmidpunct}
{\mcitedefaultendpunct}{\mcitedefaultseppunct}\relax
\EndOfBibitem
\bibitem[Monserrat \latin{et~al.}(2020)Monserrat, Brandenburg, Engel, and
  Cheng]{monserrat2020liquid}
Monserrat,~B.; Brandenburg,~J.~G.; Engel,~E.~A.; Cheng,~B. Liquid water
  contains the building blocks of diverse ice phases. \emph{Nature
  communications} \textbf{2020}, \emph{11}, 1--8\relax
\mciteBstWouldAddEndPuncttrue
\mciteSetBstMidEndSepPunct{\mcitedefaultmidpunct}
{\mcitedefaultendpunct}{\mcitedefaultseppunct}\relax
\EndOfBibitem
\bibitem[Himanen \latin{et~al.}(2020)Himanen, J{\"a}ger, Morooka, Canova,
  Ranawat, Gao, Rinke, and Foster]{himanen2020dscribe}
Himanen,~L.; J{\"a}ger,~M.~O.; Morooka,~E.~V.; Canova,~F.~F.; Ranawat,~Y.~S.;
  Gao,~D.~Z.; Rinke,~P.; Foster,~A.~S. DScribe: Library of descriptors for
  machine learning in materials science. \emph{Computer Physics Communications}
  \textbf{2020}, \emph{247}, 106949\relax
\mciteBstWouldAddEndPuncttrue
\mciteSetBstMidEndSepPunct{\mcitedefaultmidpunct}
{\mcitedefaultendpunct}{\mcitedefaultseppunct}\relax
\EndOfBibitem
\bibitem[Donkor \latin{et~al.}(0)Donkor, Laio, and
  Hassanali]{laiodonkorhassanali2023}
Donkor,~E.~D.; Laio,~A.; Hassanali,~A. Do Machine-Learning Atomic Descriptors
  and Order Parameters Tell the Same Story? The Case of Liquid Water.
  \emph{Journal of Chemical Theory and Computation} \textbf{0}, \emph{0}, null,
  PMID: 36920997\relax
\mciteBstWouldAddEndPuncttrue
\mciteSetBstMidEndSepPunct{\mcitedefaultmidpunct}
{\mcitedefaultendpunct}{\mcitedefaultseppunct}\relax
\EndOfBibitem
\bibitem[Sormani \latin{et~al.}(2019)Sormani, Rodriguez, and
  Laio]{sormani2019explicit}
Sormani,~G.; Rodriguez,~A.; Laio,~A. Explicit Characterization of the
  Free-Energy Landscape of a Protein in the Space of All Its C$\alpha$ Carbons.
  \emph{Journal of chemical theory and computation} \textbf{2019}, \emph{16},
  80--87\relax
\mciteBstWouldAddEndPuncttrue
\mciteSetBstMidEndSepPunct{\mcitedefaultmidpunct}
{\mcitedefaultendpunct}{\mcitedefaultseppunct}\relax
\EndOfBibitem
\bibitem[d’Errico \latin{et~al.}(2021)d’Errico, Facco, Laio, and
  Rodriguez]{DERRICO2021476}
d’Errico,~M.; Facco,~E.; Laio,~A.; Rodriguez,~A. Automatic topography of
  high-dimensional data sets by non-parametric density peak clustering.
  \emph{Information Sciences} \textbf{2021}, \emph{560}, 476--492\relax
\mciteBstWouldAddEndPuncttrue
\mciteSetBstMidEndSepPunct{\mcitedefaultmidpunct}
{\mcitedefaultendpunct}{\mcitedefaultseppunct}\relax
\EndOfBibitem
\bibitem[Diaz-Papkovich \latin{et~al.}(2019)Diaz-Papkovich,
  Anderson-Trocm{\'e}, Ben-Eghan, and Gravel]{diaz2019umap}
Diaz-Papkovich,~A.; Anderson-Trocm{\'e},~L.; Ben-Eghan,~C.; Gravel,~S. UMAP
  reveals cryptic population structure and phenotype heterogeneity in large
  genomic cohorts. \emph{PLoS genetics} \textbf{2019}, \emph{15},
  e1008432\relax
\mciteBstWouldAddEndPuncttrue
\mciteSetBstMidEndSepPunct{\mcitedefaultmidpunct}
{\mcitedefaultendpunct}{\mcitedefaultseppunct}\relax
\EndOfBibitem
\bibitem[van~der Maaten(2014)]{tsne}
van~der Maaten,~L. Accelerating t-SNE using Tree-Based Algorithms.
  \emph{Journal of Machine Learning Research} \textbf{2014}, \emph{15},
  3221--3245\relax
\mciteBstWouldAddEndPuncttrue
\mciteSetBstMidEndSepPunct{\mcitedefaultmidpunct}
{\mcitedefaultendpunct}{\mcitedefaultseppunct}\relax
\EndOfBibitem
\bibitem[Sprik(2000)]{SPRIK2000139}
Sprik,~M. Computation of the pK of liquid water using coordination constraints.
  \emph{Chemical Physics} \textbf{2000}, \emph{258}, 139--150\relax
\mciteBstWouldAddEndPuncttrue
\mciteSetBstMidEndSepPunct{\mcitedefaultmidpunct}
{\mcitedefaultendpunct}{\mcitedefaultseppunct}\relax
\EndOfBibitem
\bibitem[Gasparotto \latin{et~al.}(2016)Gasparotto, Hassanali, and
  Ceriotti]{GasparottoHassanaliCeriotti2016}
Gasparotto,~P.; Hassanali,~A.~A.; Ceriotti,~M. Probing Defects and Correlations
  in the Hydrogen-Bond Network of ab Initio Water. \emph{Journal of Chemical
  Theory and Computation} \textbf{2016}, \emph{12}, 1953--1964, PMID:
  26881726\relax
\mciteBstWouldAddEndPuncttrue
\mciteSetBstMidEndSepPunct{\mcitedefaultmidpunct}
{\mcitedefaultendpunct}{\mcitedefaultseppunct}\relax
\EndOfBibitem
\bibitem[Henchman and Irudayam(2010)Henchman, and
  Irudayam]{henchman2010topological}
Henchman,~R.~H.; Irudayam,~S.~J. Topological hydrogen-bond definition to
  characterize the structure and dynamics of liquid water. \emph{The Journal of
  Physical Chemistry B} \textbf{2010}, \emph{114}, 16792--16810\relax
\mciteBstWouldAddEndPuncttrue
\mciteSetBstMidEndSepPunct{\mcitedefaultmidpunct}
{\mcitedefaultendpunct}{\mcitedefaultseppunct}\relax
\EndOfBibitem
\bibitem[Markovitch \latin{et~al.}(2008)Markovitch, Chen, Izvekov, Paesani,
  Voth, and Agmon]{markovitch2008special}
Markovitch,~O.; Chen,~H.; Izvekov,~S.; Paesani,~F.; Voth,~G.~A.; Agmon,~N.
  Special pair dance and partner selection: Elementary steps in proton
  transport in liquid water. \emph{The Journal of Physical Chemistry B}
  \textbf{2008}, \emph{112}, 9456--9466\relax
\mciteBstWouldAddEndPuncttrue
\mciteSetBstMidEndSepPunct{\mcitedefaultmidpunct}
{\mcitedefaultendpunct}{\mcitedefaultseppunct}\relax
\EndOfBibitem
\bibitem[Kulig and Agmon(2014)Kulig, and Agmon]{KuligAgmon2014}
Kulig,~W.; Agmon,~N. Both Zundel and Eigen Isomers Contribute to the IR
  Spectrum of the Gas-Phase H9O4+ Cluster. \emph{The Journal of Physical
  Chemistry B} \textbf{2014}, \emph{118}, 278--286, PMID: 24344636\relax
\mciteBstWouldAddEndPuncttrue
\mciteSetBstMidEndSepPunct{\mcitedefaultmidpunct}
{\mcitedefaultendpunct}{\mcitedefaultseppunct}\relax
\EndOfBibitem
\bibitem[Fournier \latin{et~al.}(2014)Fournier, Johnson, Wolke, Weddle, Wolk,
  and Johnson]{johnson2014}
Fournier,~J.~A.; Johnson,~C.~J.; Wolke,~C.~T.; Weddle,~G.~H.; Wolk,~A.~B.;
  Johnson,~M.~A. Vibrational spectral signature of the proton defect in the
  three-dimensional $H^+(H_2O)_{21}$ cluster. \emph{Science} \textbf{2014},
  \emph{344}, 1009--1012\relax
\mciteBstWouldAddEndPuncttrue
\mciteSetBstMidEndSepPunct{\mcitedefaultmidpunct}
{\mcitedefaultendpunct}{\mcitedefaultseppunct}\relax
\EndOfBibitem
\bibitem[Hassanali \latin{et~al.}(2011)Hassanali, Prakash, Eshet, and
  Parrinello]{HassanaliPrakashEshetParrinello2011}
Hassanali,~A.; Prakash,~M.~K.; Eshet,~H.; Parrinello,~M. On the recombination
  of hydronium and hydroxide ions in water. \emph{Proceedings of the National
  Academy of Sciences} \textbf{2011}, \emph{108}, 20410--20415\relax
\mciteBstWouldAddEndPuncttrue
\mciteSetBstMidEndSepPunct{\mcitedefaultmidpunct}
{\mcitedefaultendpunct}{\mcitedefaultseppunct}\relax
\EndOfBibitem
\bibitem[Bj\"{o}rneholm \latin{et~al.}(2016)Bj\"{o}rneholm, Hansen, Hodgson,
  and Liu]{chemrevinterfaces2016}
Bj\"{o}rneholm,~O.; Hansen,~M.~H.; Hodgson,~A.; Liu,~L. Water at Interfaces.
  \emph{Chemical Reviews} \textbf{2016}, \emph{116}, 7698--7726, PMID:
  27232062\relax
\mciteBstWouldAddEndPuncttrue
\mciteSetBstMidEndSepPunct{\mcitedefaultmidpunct}
{\mcitedefaultendpunct}{\mcitedefaultseppunct}\relax
\EndOfBibitem
\bibitem[Creazzo \latin{et~al.}(2020)Creazzo, Pezzotti, Bougueroua, Serva,
  Sponer, Saija, Cassone, and Gaigeot]{creazzo2020}
Creazzo,~F.; Pezzotti,~S.; Bougueroua,~S.; Serva,~A.; Sponer,~J.; Saija,~F.;
  Cassone,~G.; Gaigeot,~M.-P. Enhanced conductivity of water at the electrified
  air–water interface: a DFT-MD characterization. \emph{Phys. Chem. Chem.
  Phys.} \textbf{2020}, \emph{22}, 10438--10446\relax
\mciteBstWouldAddEndPuncttrue
\mciteSetBstMidEndSepPunct{\mcitedefaultmidpunct}
{\mcitedefaultendpunct}{\mcitedefaultseppunct}\relax
\EndOfBibitem
\bibitem[Di~Pino \latin{et~al.}()Di~Pino, Perez~Sirkin, Morzan, Sánchez,
  Hassanali, and Scherlis]{solana23}
Di~Pino,~S.; Perez~Sirkin,~Y.~A.; Morzan,~U.~N.; Sánchez,~V.~M.;
  Hassanali,~A.; Scherlis,~D.~A. Water Self-Dissociation is Insensitive to
  Nanoscale Environments. \emph{Angewandte Chemie International Edition}
  \emph{n/a}, e202306526\relax
\mciteBstWouldAddEndPuncttrue
\mciteSetBstMidEndSepPunct{\mcitedefaultmidpunct}
{\mcitedefaultendpunct}{\mcitedefaultseppunct}\relax
\EndOfBibitem
\bibitem[Tielrooij \latin{et~al.}(2009)Tielrooij, Cox, and Bakker]{bakker2009}
Tielrooij,~K.~J.; Cox,~M.~J.; Bakker,~H.~J. Effect of Confinement on
  Proton-Transfer Reactions in Water Nanopools. \emph{ChemPhysChem}
  \textbf{2009}, \emph{10}, 245--251\relax
\mciteBstWouldAddEndPuncttrue
\mciteSetBstMidEndSepPunct{\mcitedefaultmidpunct}
{\mcitedefaultendpunct}{\mcitedefaultseppunct}\relax
\EndOfBibitem
\bibitem[Saitta \latin{et~al.}(2012)Saitta, Saija, and Giaquinta]{saitta2012}
Saitta,~A.~M.; Saija,~F.; Giaquinta,~P.~V. Ab Initio Molecular Dynamics Study
  of Dissociation of Water under an Electric Field. \emph{Phys. Rev. Lett.}
  \textbf{2012}, \emph{108}, 207801\relax
\mciteBstWouldAddEndPuncttrue
\mciteSetBstMidEndSepPunct{\mcitedefaultmidpunct}
{\mcitedefaultendpunct}{\mcitedefaultseppunct}\relax
\EndOfBibitem
\bibitem[Cassone \latin{et~al.}(2014)Cassone, Giaquinta, Saija, and
  Saitta]{cassonejpcb14}
Cassone,~G.; Giaquinta,~P.~V.; Saija,~F.; Saitta,~A.~M. Proton Conduction in
  Water Ices under an Electric Field. \emph{The Journal of Physical Chemistry
  B} \textbf{2014}, \emph{118}, 4419--4424, PMID: 24689531\relax
\mciteBstWouldAddEndPuncttrue
\mciteSetBstMidEndSepPunct{\mcitedefaultmidpunct}
{\mcitedefaultendpunct}{\mcitedefaultseppunct}\relax
\EndOfBibitem
\bibitem[Sun \latin{et~al.}(2015)Sun, Ruzsinszky, and Perdew]{SCAN}
Sun,~J.; Ruzsinszky,~A.; Perdew,~J.~P. Strongly Constrained and Appropriately
  Normed Semilocal Density Functional. \emph{Phys. Rev. Lett.} \textbf{2015},
  \emph{115}, 036402\relax
\mciteBstWouldAddEndPuncttrue
\mciteSetBstMidEndSepPunct{\mcitedefaultmidpunct}
{\mcitedefaultendpunct}{\mcitedefaultseppunct}\relax
\EndOfBibitem
\end{mcitethebibliography}

\clearpage

\section{Supporting Information}

\begin{figure}[h]
\centering
\includegraphics[width=1.0\textwidth]{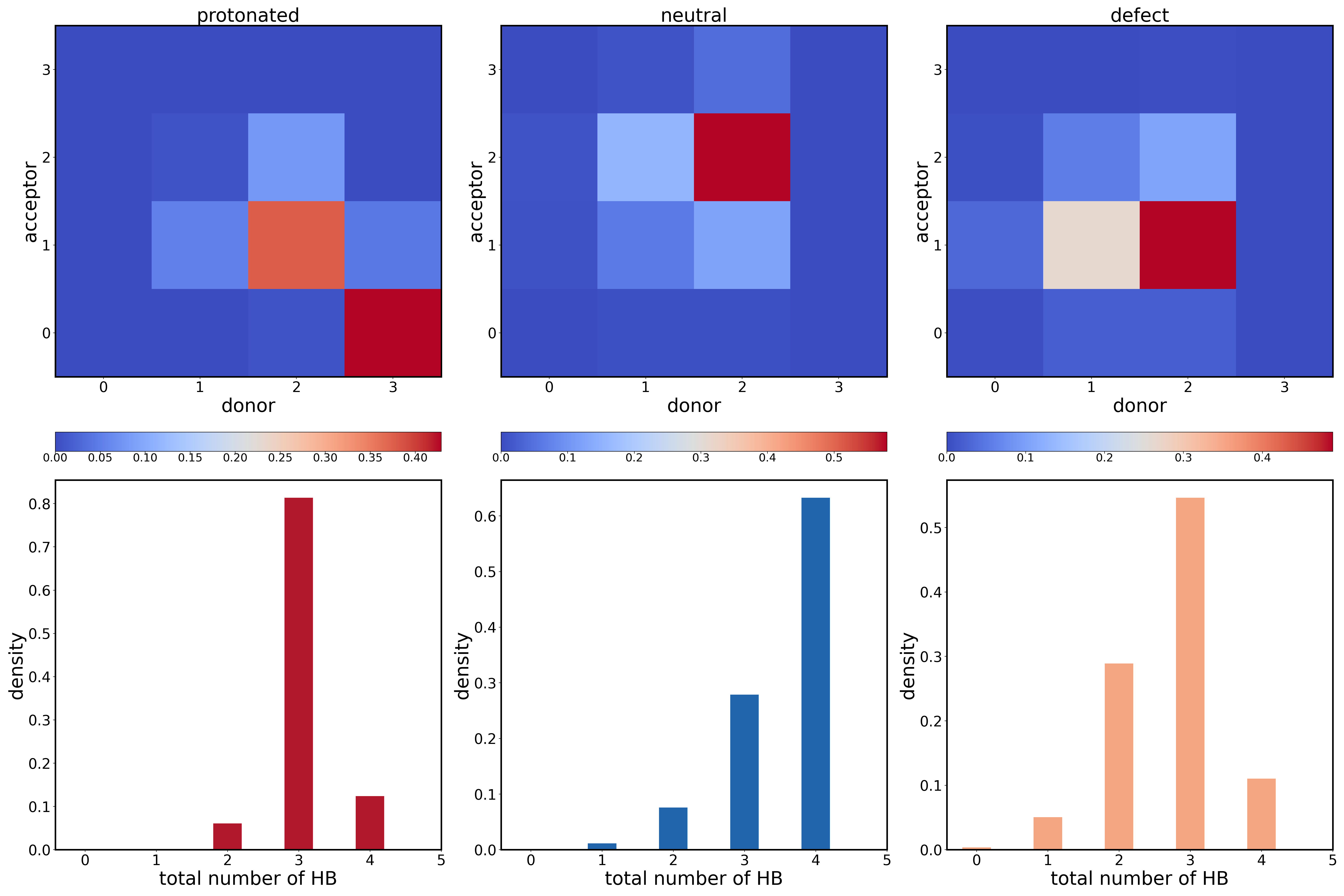}
\caption{Hydrogen bonding coordination patterns for the three different clusters obtained with $Z=2.5$ (protonated: left panels, neutral: middle panels, neutral defect: right panels). Top panels show the two-dimensional (2D) distributions for the concentrations of the number of accepting/donating hydrogen bonds that each cluster participates in. Bottom panel displays instead the distributions of the total number of hydrogen bonds obtained for the different clusters.}
\label{sifig2}
\end{figure}

\begin{figure}[h]
\centering
\includegraphics[width=1.0\textwidth]{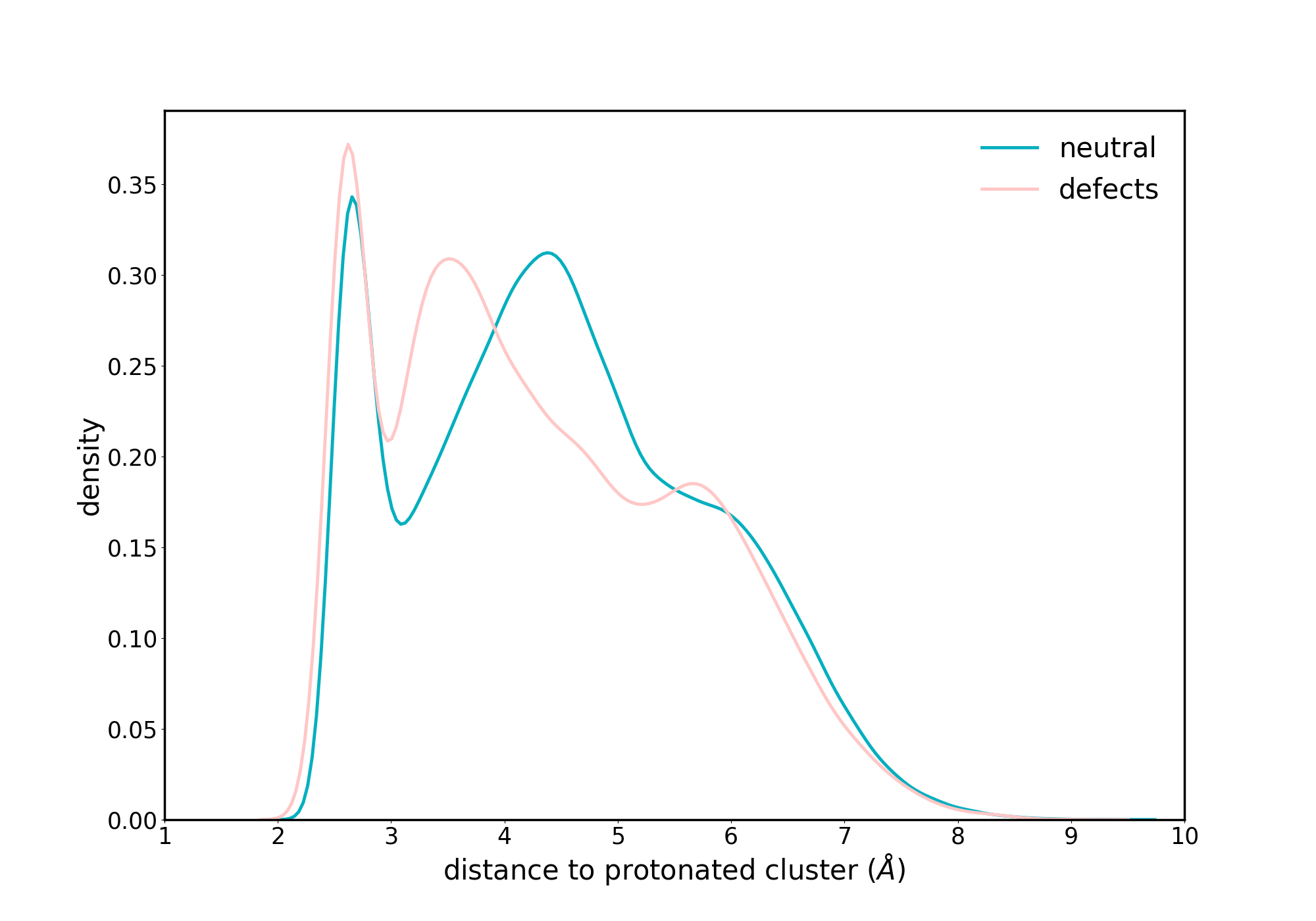}
\caption{Pair-correlation function determined between the oxygen atom of the protonated cluster with respect to the neutral (blue) and the neutral defect (pink) clusters. We observe that, on average, the neutral defect cluster has a slightly larger tendency to form closer to the protonated cluster.}
\label{sifig3}
\end{figure}

\clearpage

\begin{figure}[h]
\centering
\includegraphics[width=1.0\textwidth]{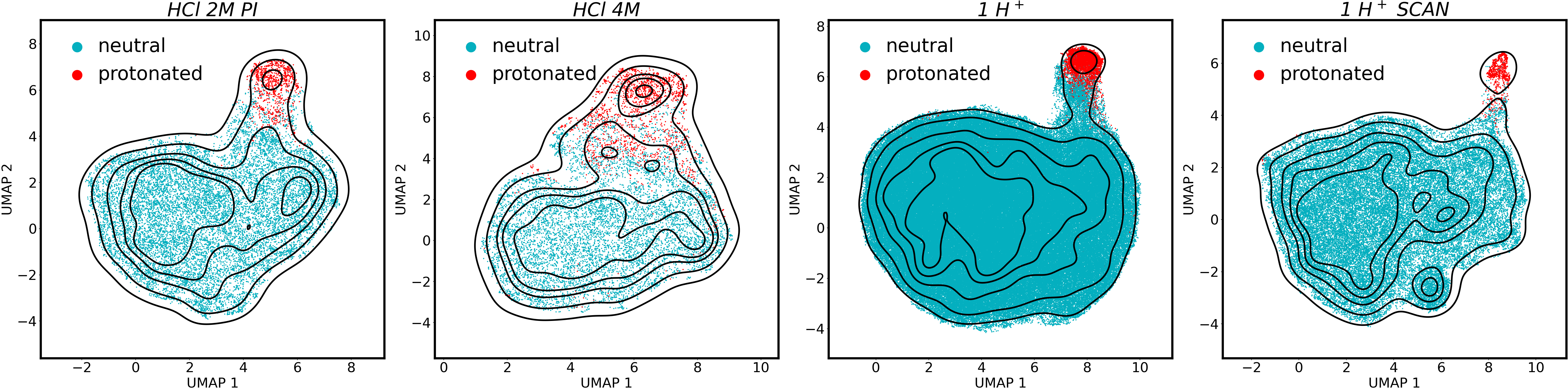}
\caption{Comparison of the clustering outcomes obtained with $Z=3.0$ for the 2~M HCl from path integral (PI) \emph{ab initio} molecular dynamics (AIMD) including Nuclear Quantum Effects, the 4~M HCl from AIMD with classical nuclei, AIMD of a single proton in water (1H$^+$), and a single proton in water using the SCAN functional (1H$^+$-SCAN)\cite{SCAN}. Similar to the HCl 4~M case, we used a $Z=2.0$ value due to limited statistics. Overall, we observe that our protocol yields the same conclusions, independently from the inclusion of Nuclear Quantum Effects as well as using different qualities of electronic structure when treating the exchange correlation functional.}
\label{sifig1}
\end{figure}

\begin{figure}[h]
\centering
\includegraphics[width=1.0\textwidth]{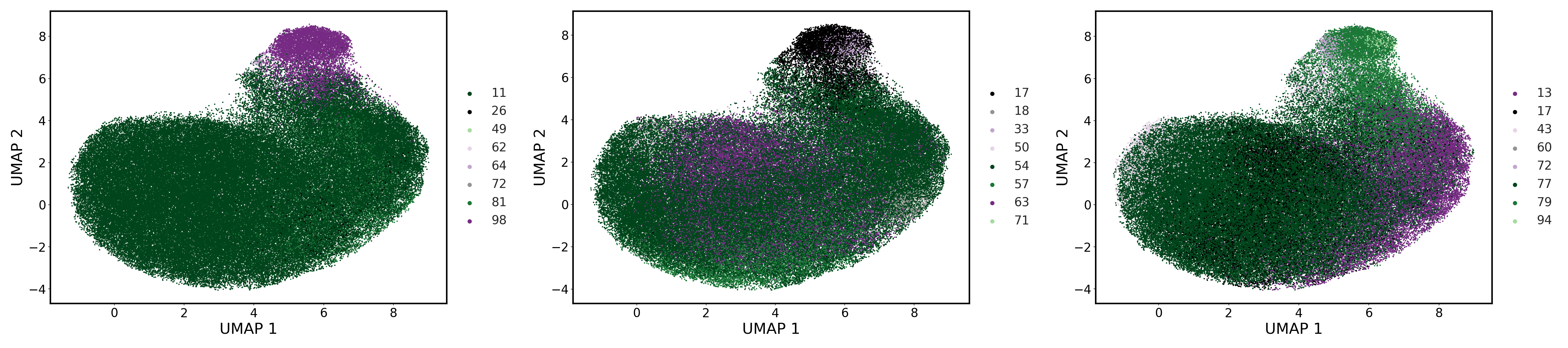}
\caption{In the main text we stated that reduction of the Z parameter to 2.0 can lead to splitting of the protonated cluster. Besides the fact that these clusters do not partition into Eigen or Zundel states, we also note that these features are not statistically significant. The Figure shows the UMAP projections obtained from randomly sampling 3 datasets of half the size from the original one where we observe very different partitioning implying that the observed splitting arises from statistical fluctuations.}
\label{sifig4}
\end{figure}

\clearpage

\section{Selection of SOAP Parameters for Milestone Structures}

As discussed in the main text, to compare the proximity of the putative protonated atomic environments in HCl to the Eigen and Zundel species, respectively, we identified two milestone topologies corresponding to these two molecular structures. In particular, the Eigen and Zundel motifs are well defined in gas-phase -- low-temperature -- clusters. We thus constructed SOAP descriptors based on these structures that directly probe the Eigen and Zundel states. For the Eigen milestone, we used the oxygen atom as the reference center, using the same $\sigma$ as before, $n_{max}=10$ and $l_{max}=6$ with a radial cutoff of $2.15$~{\AA}. On the other hand, the shared proton has been used as the reference center for the Zundel milestone structure, with the same $\sigma$, $n_{max}$, $l_{max}$ as in the Eigen case but with a larger radial cutoff of $2.8$~{\AA}. The shared proton is identified by choosing the closest oxygen atom to the one associated with the protonated cluster (1NN in the main text) and then selecting the hydrogen atom that is sandwiched between them. The two different cutoffs ensure that for the Eigen cation we probe an environment representative of the (H$_3$O)$^+$ moiety whilst for the Zundel cation that of the (H$_{5}$O$_{2}$)$^+$ complex.

\begin{figure}[h]
\centering
\includegraphics[width=1.0\textwidth]{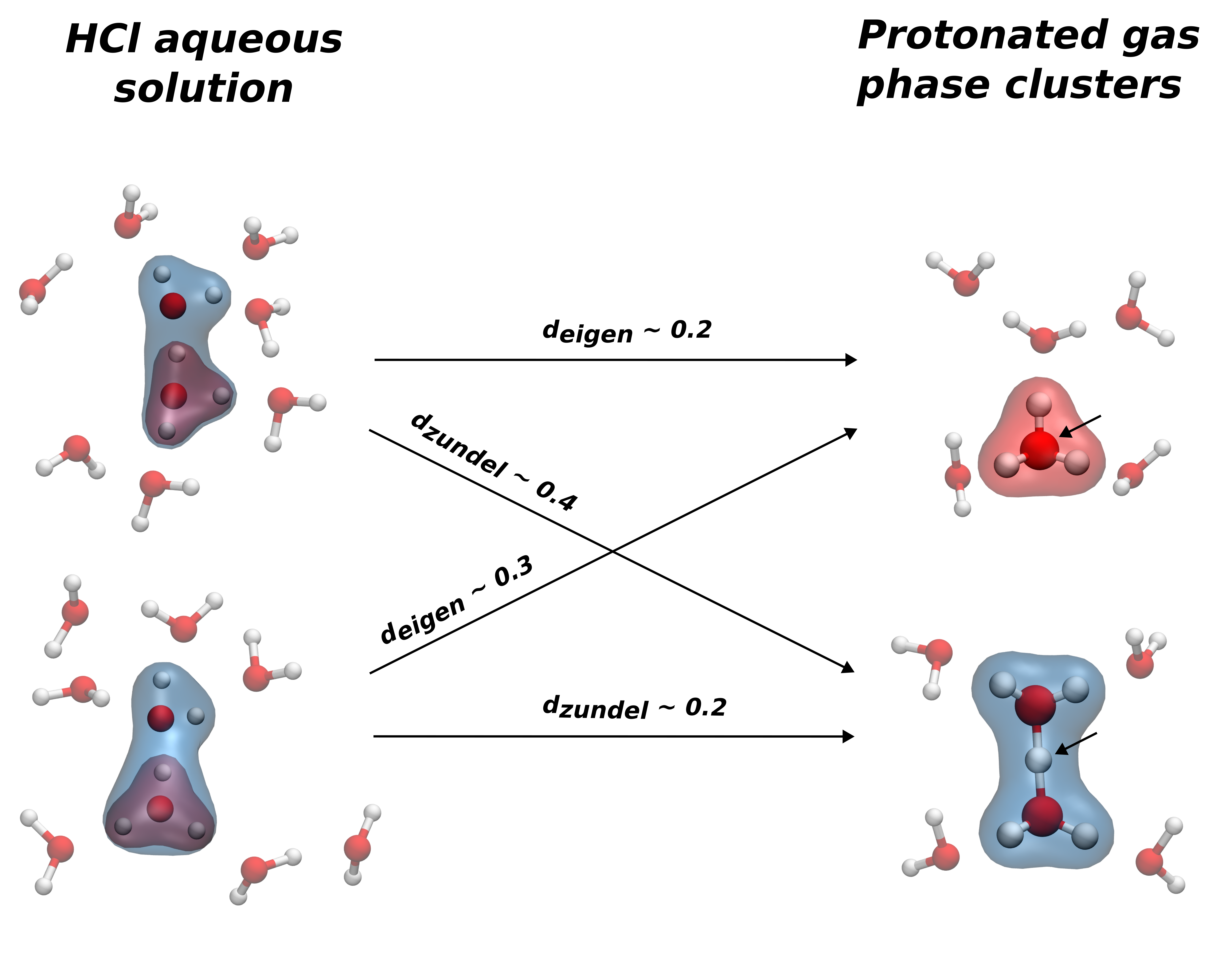}
\caption{Schematic of the post processing procedure used to re calculate the SOAP descriptors constraining the environment to the basic structures of Eigen ($H_3O^+$) and Zundel ($H_5O_2^+$) ions. Configurations taken from our HCl trajectories (left) are compared to idealized configurations of Eigen and Zundel ions taken from protonated gas phase clusters (right). Note that SOAP descriptors are built using oxygen atoms as centers for the comparison to Eigen milestones but hydrogen atoms are used as centers for the case of Zundel milestones, as indicated by the black arrows in the figure.}
\label{sifig6}
\end{figure}

\end{document}